\documentclass[iop,revtex4]{emulateapj}

\usepackage{float}
\usepackage{amsmath}
\usepackage{array}

\newcommand{\mytilde}{\raise.19ex\hbox{$\scriptstyle\sim$}}

\slugcomment{}

\shorttitle{Weak-lensing Study of PLCK~G287.0+32.9}
\shortauthors{Finner et al. }

\begin{document}

\title{MC$^2$: Subaru and {\it Hubble Space Telescope} Weak-lensing Analysis of \\ the Double Radio Relic Galaxy Cluster PLCK~G287.0+32.9}

\author{Kyle Finner\altaffilmark{1}, M. James Jee\altaffilmark{1,2}, Nathan Golovich\altaffilmark{2,3}, David Wittman\altaffilmark{2}, William Dawson\altaffilmark{3}, Daniel Gruen\altaffilmark{4,5,6}, \\ Anton M. Koekemoer\altaffilmark{7}, Brian C. Lemaux\altaffilmark{2}, and Stella Seitz\altaffilmark{8,9}}
\altaffiltext{1}{Department of Astronomy, Yonsei University, 50 Yonsei-ro, Seoul 03722, Korea; kylefinner@gmail.com, 
mkjee@yonsei.ac.kr}
\altaffiltext{2}{Department of Physics, University of California, Davis, One Shields Avenue, Davis, CA 95616, USA}
\altaffiltext{3}{Lawrence Livermore National Laboratory, P.O. Box 808 L-210, Livermore, CA, 94551, USA}
\altaffiltext{4}{SLAC National Accelerator Laboratory, Menlo Park, CA 94025, USA}
\altaffiltext{5}{KIPAC, Physics Department, Stanford University, Stanford, CA 94305, USA}
\altaffiltext{6}{Einstein Fellow}
\altaffiltext{7}{Space Telescope Science Institute, 3700 San Martin Dr., Baltimore, MD, 21218, USA}
\altaffiltext{8}{University Observatory Munich, Scheinerstrasse 1, D-81679 Munich, Germany}
\altaffiltext{9}{Max Planck Institute for Extraterrestrial Physics, Giessenbachstrasse, D-85741 Garching, Germany}

\begin{abstract}

The second most significant detection of the Planck Sunyaev Zel'dovich survey, PLCK~G287.0+32.9 ($z=0.385$) boasts two similarly bright radio relics and a radio halo. One radio relic is located $\mytilde 400$ kpc northwest of the X-ray peak and the other $\mytilde 2.8$ Mpc to the southeast. This large difference suggests that a complex merging scenario is required. A key missing puzzle for the merging scenario reconstruction is the underlying dark matter distribution in high resolution.
We present a joint Subaru Telescope and {\it Hubble Space Telescope} weak-lensing analysis of the cluster. Our analysis shows that the mass distribution features four significant substructures. Of the substructures, a primary cluster of mass $M_{200\text{c}}=1.59^{+0.25}_{-0.22}\times 10^{15} \ h^{-1}_{70} \ \text{M}_{\odot}$ dominates the weak-lensing signal. This cluster is likely to be undergoing a merger with one (or more) subcluster whose mass is approximately a factor of 10 lower. One candidate is the subcluster
of mass $M_{200\text{c}}=1.16^{+0.15}_{-0.13}\times 10^{14} \ h^{-1}_{70} \ \text{M}_{\odot}$ located $\mytilde 400$ kpc to the southeast. The location of this subcluster suggests that its interaction with the primary cluster could be the source of the NW radio relic. Another subcluster is detected $\mytilde 2$ Mpc to the SE of the X-ray peak with mass $M_{200\text{c}}=1.68^{+0.22}_{-0.20}\times 10^{14} \ h^{-1}_{70} \ \text{M}_{\odot}$. This SE subcluster is in the vicinity of the SE radio relic and may have created the SE radio relic during a past merger with the primary cluster. The fourth subcluster, $M_{200\text{c}}=1.87^{+0.24}_{-0.22}\times 10^{14} \ h^{-1}_{70} \ \text{M}_{\odot}$, is northwest of the X-ray peak and beyond the NW radio relic. 

\end{abstract}


\section{Introduction}
Among the largest of the gravitationally bound structures, galaxy clusters are comprised of about 90\% dark matter, 9\% plasma intra-cluster medium (ICM), and 1\% stars. Galaxy clusters grow by the accretion of gas, galaxies, and other clusters. The most energetic of these events is a collision of two massive galaxy clusters, releasing $\mytilde 10^{64}\ \text{erg}$ of energy \citep{sarazin2002}. 

During a cluster merger, galaxies and dark matter interact predominantly through the gravitational force and for the most part can be considered as collisionless. In contrast, the gas particles of the ICM are collisional and can suffer an abrupt dissipation of energy from ram-pressure. The difference in the two interaction mechanisms has been observed as a lag of the plasma ICM while the galaxies continue along their effectively unchanged orbits. This is apparent in ``the Bullet Cluster'' (1E 0657-558) where two subclusters of galaxies are clearly separated and the X-ray emitting ICM spans the space between \citep{clowe2006bullet}.

Within the ICM of highly energetic galaxy cluster mergers, shock waves have been observed. As the two clusters collide, the supersonic velocity ($\mytilde 2,000 - 3,000$ km s$^{-1}$) of the collision causes a rapid compression of gas at the intersection of the cores producing a shock. Shocks observed in X-ray emissions appear in low-surface brightness regions of the ICM with a surface brightness profile showing a discontinuous jump perpendicular to the shock front. Numerical simulations have reproduced shocks ranging from a few 100 kpc to a few Mpc in size \citep{ryu2003}. Even though the shocks are large, their detection by X-ray emission is difficult because of low surface brightness.

In the low density outskirts of merging galaxy clusters, diffuse radio emissions from the ICM, known as radio relics, are believed to trace shocks. To date, approximately 50 clusters have been detected with radio relics. For a recent review of diffuse radio emissions from galaxy clusters see \cite{feretti2012}. Among the clusters exhibiting radio relics, 18 have double radio relics. The systems hosting radio relics are believed to arise in clusters that have a merging axis near the plane of the sky  \citep{golovich2017}. Diffuse radio emissions have also been observed in the centers of about 40 galaxy clusters \citep{feretti2012}. In this region, the radio emission is called a radio halo and it is believed to be synchrotron emission from the turbulent ICM caused by a major merger.

The Merging Cluster Collaboration (MC$^2$) has selected a sample of 29 merging clusters with elongated radio relics in an effort to improve our understanding of galaxy cluster physics and to constrain dark matter properties. This sample contains three galaxy clusters with a visible radio halo and double radio relics confirmed: RXC J1314.4-2515, CIZA J2242.8+5301, and PLCK~G287.0+32.9. Of these, PLCK~G287.0+32.9 is the most massive and highest redshift and is the focus of this paper.

PLCK~G287.0+32.9 was first detected by the Planck telescope through the Sunyaev Zel'dovich (SZ) effect. It was reported as the second most significant detection in the Planck collaboration all-sky early Sunyaev Zel'dovich cluster sample \citep{planck2011a}. Warranted by its high SZ S/N, PLCK~G287.0+32.9 was included in \textit{XMM-Newton} follow-up observations of significant SZ detections \citep{planck2011b}. \textit{XMM-Newton} measurements found the temperature of the intracluster medium to be $12.86\pm0.42\ \mathrm{keV}$. The mass of the cluster within $R_{500\text{c}}$ was inferred to be $M_{500\text{c}}=15.72\pm0.27\times10^{14}\ h^{-1}_{70}\ \text{M}_{\odot}$ (only statistical error considered) from the mass-Compton Y parameter relation. The X-ray emissions were further analyzed by \cite{bagchi2011} and found to have a disturbed morphology that is compressed in the north. 

In addition to the X-ray analysis, \cite{bagchi2011} performed the first radio investigation of the cluster. Equipped with 150 MHz Giant Meterwave Radio Telescope (GMRT) and 1.4 GHz Very Large Array (VLA) radio data, the discovery of two mega-parsec sized radio relics was made, one north and the other south of the cluster. \cite{bagchi2011} estimated the projected separation of the relics to be $\mytilde4.4$ Mpc. A third diffuse radio feature closer to the X-ray peak (named Y) was discussed but, with poor resolution imaging, it was unclear whether this feature was a relic on the periphery of the cluster or central and part of a radio halo. In conjunction with the X-ray observations, the radio features distinguished PLCK~G287.0+32.9 as a merging galaxy cluster.

A clearer picture of the radio emissions from PLCK~G287.0+32.9 was provided by \cite{bonafede2014} with more sensitive (5-7 times) and higher resolution (4-8 times) GMRT and VLA radio observations. They determined that the previously named northern radio relic is in fact  emission from a radio galaxy and that the poorly resolved emissions near the cluster center are indeed a radio relic. This realization decreases the projected separation of the radio relics to $\mytilde 3.2$ Mpc with the northern relic $\mytilde0.4$ Mpc and the southern relic $\mytilde2.8$ Mpc from the X-ray peak. Since the newly coined relic is much closer to the X-ray peak, the asymmetric distances require a  complex merging scenario. \cite{bonafede2014} also confirmed the existence of a radio halo approximately 1.3 Mpc in diameter. The spectral index of the southern relic decreases with distance from the cluster as expected in diffuse synchrotron aging. However, interestingly, the spectral index of the northern relic decreases initially and then increases sharply with distance from the cluster. This reverse trend is hard to reconcile with the conventional understanding of the Diffusive Shock Acceleration (DSA) spectral aging mechanism \citep{bonafede2014}.

With a complex merging scenario unfolding from the X-ray and radio observations, a considerable improvement in our understanding of the merger history is possible if we know the underlying dark matter distribution. Mass estimations that rely on equilibrium conditions are inadequate approximations for merging galaxy clusters. The first mass estimation without a hydrostatic equilibrium assumption for 
PLCK~G287.0+32.9 was included in a weak-lensing analysis of SZ-selected clusters \citep{gruen2014weak} with MPG/ESO telescope/Wide-Field Imager (WFI) observations. Based on a single NFW halo fit, centered at the brightest cluster galaxy (BCG), the mass was found to be $M_{200\text{m}}=3.77^{+0.95}_{-0.76}\times10^{15}\ h^{-1}_{70}\ \text{M}_{\odot}$. Additional support for this extremely large mass is provided in a recent strong-lensing analysis by \cite{zitrin2017} where they identified 20 sets of multiply-imaged galaxies in \textit{Hubble Space Telescope (HST)} data and determined the cluster to have an effective Einstein radius of $\theta_\text{E}\simeq42\arcsec$ (when they assumed $z_s\simeq2$).  

In this study, we present the first constraints on the mass distribution of PLCK~G287.0+32.9 from a weak-lensing analysis of Subaru and {\it HST} observations. Our analysis with this new data set provides substantial improvements over the previous weak-lensing study of \cite{gruen2014weak}. In terms of the source density, our deeper Subaru imaging data alone yields more than four times the background galaxies usable for lensing analysis. This enhancement enable us to achieve much tighter constraints on the cluster mass and to resolve merging substructures. In addition, the high-resolution \textit{HST} coverage in the central $3\arcmin \times 3\arcmin$ region allows us to obtain more than 100 galaxies arcmin$^{-2}$, which is essential to reveal substructures in the densest region. 

Our paper is organized in the following manner. In Section 2, we describe our data reduction techniques for Subaru and {\it HST} photometric data and Keck spectroscopic data. Modeling of the Point Spread Function is explained in Section 3, followed by a brief overview of weak-lensing theory in Section 4. We describe our shape measurement in Section 5 and the resulting source catalog creation in Section 6. Section 7 briefly explains the redshift estimation for the source catalogs. In Section 8 we present our results. We then discuss the results in Section 9 and conclude in Section 10.

Throughout this work, we assume a flat $\Lambda$CDM cosmology with $H_0=70$ km s$^{-1}$ Mpc$^{-1}$, $\Omega_m=0.3$, and $\Omega_\Lambda=0.7$. At the redshift of PLCK~G287.0+32.9 ($z=0.385$), $1\arcsec$ corresponds to 5.25 kpc. We use the notation of $R_{\Delta_c}$ and $R_{\Delta_m}$ to indicate the radius of a sphere inside which the mean matter density is $\Delta$ times the critical matter density (c) or the mean matter density (m) of the Universe at the cluster redshift. The analogous notation is used to present mass enclosed at the defined radii. The factor $h^{-1}_{70} = H_0 / (100\ \text{km}\ \text{s}^{-1}\ \text{Mpc}^{-1})$ is included in masses derived in this work. All magnitudes are defined in the AB magnitude system.

\section{Observations}
\subsection{Subaru/SuprimeCam Data Reduction}
Observations of the galaxy cluster PLCK~G287.0+32.9 were carried out with the 8.2m Subaru Telescope on Mauna Kea, the night of 2014 February 26 (PI: D. Wittman). We obtained seven pointings of the $g$-band filter with total integration time of 753s and ten pointings of the $r$-band filter with total integration time of 2,913s. The median seeing for $g$ and $r$ were 1\farcs17 and 1\farcs13, respectively. In our analysis, we use the $r$-band image to measure the shapes of galaxies (Section \ref{section:shape measurement}) and use $g-r$ color to select galaxies (Section \ref{section:source selection}).

Weak-lensing analyses can be hampered by spurious features such as bad columns, cosmic rays, bleeding trails, and diffraction spikes. A rotation of $30^\circ$ for $g$ and $15^\circ$ for $r$ was introduced between pointings to minimize the effect of these features. This method has been shown to increase the number of usable galaxies available for weak-lensing studies \citep{jee2015sausage, jee2016toothbrush}, minimizing the number of galaxies contaminated by the bleeding trails of saturated stars. The SuprimeCam's $34\arcmin \times 27\arcmin $ field of view allowed each pointing to amply cover the galaxy cluster region ($\mytilde 11$ Mpc $\times\ 8.5$ Mpc at z=0.385). 

Special care is required during data reduction to produce a weak-lensing quality image. The Subaru supplied \texttt{SDFRED2} package\footnote{http://subarutelescope.org/Observing/Instruments/SCam/sdfred} is used for the basic steps: over-scan subtraction, bias correction, and flat-fielding. \texttt{SDFRED2} is also used to correct for the geometric distortion due to the telescope optics and distortion due to the differential atmospheric dispersion. Each frame is then passed through \texttt{Source Extractor} \citep{1996bertin} to prepare a catalog of objects to be used in \texttt{SCAMP} \citep{2006bertin}. \texttt{SCAMP} is applied to correct for residual distortions and to refine the World Coordinate System (WCS) for each frame. \texttt{SCAMP} uses a reference catalog to compute astrometry. A quality reference catalog is required to achieve the precision needed for weak-lensing analysis. We use the Two Micron All Sky Survey \citep[][2MASS hereafter]{20062mass} catalog as reference because it has been shown \citep{2012heymans, jee2015sausage, jee2016toothbrush} that it provides the most accurate astrometric solution (measured rms $\mytilde$ 0.02 pixels) in our field. Photometric solutions are obtained by comparing common astronomical objects.

The final step in creating a weak-lensing quality image is to precisely co-add the frames into a mosaic image. A mean stack yields the highest S/N results. However, the frames are prone to cosmic rays and saturation trails. For this reason, the frames are co-added in two steps. First, the software \texttt{SWarp} \citep{2002bertin} is applied to the frames to create a median-stacked mosaic image. As part of the \texttt{SWarp} process, frames are translated, rotated, and distortion corrected in preparation for co-addition. These individual RESAMP frames are useful for our analysis and we store them for later use. After resampling, the frames are carefully aligned using the WCS solution from \texttt{SCAMP} and co-added by \texttt{SWarp} into a median-stacked mosaic image. The median-stacking process outputs a weight file for each component frame that contains the Gaussian variance of each pixel. The median-stacked mosaic image is then compared to the stored RESAMP frames and pixels from the RESAMP frames that deviate more than 3$\sigma$ from the median-stacked image are set to zero in the corresponding weight file. Finally, the frames are co-added in \texttt{SWarp} by weight-averaging the input frames and a mean-stacked mosaic image is created.


\subsection{Hubble Space Telescope/Advanced Camera for Surveys Data Reduction}
\textit{HST} observations of PLCK~G287.0+32.9 were made under program 14165 on 2016 August 3 (PI: S. Seitz) and program 14096 on 2017 February 21 (PI: D. Coe). Program 14165 observed the cluster using the Advanced Camera for Surveys (ACS) filters F475W, F606W, and F814W with exposure times of 2,160s (one orbit), 2,320s (one orbit), and 4,680s (two orbits), respectively. In addition, four orbits were used to image the cluster with the WFC3 filter F110W totaling 10,447s of integration. For each orbit, four exposures were taken with a half-pixel dithering pattern administered. Program 14096 carried out observations with the ACS filter F435W (2,125s of integration) and the WFC3 filters F105W (1,362s), F120W (711s), F140W (712s), and F160W (1,962s). The WFC3 observations are limited to the inner $\mytilde 2\arcmin$ around the BCG and have a larger PSF than the ACS observations. The ACS observations cover $\mytilde3\arcmin \times 3\arcmin$ central region of the cluster. For these reasons, only the four ACS filters from the two programs are used in this work. Weak-lensing shapes are measured in the ACS F814W image because of its high S/N. Cluster members are selected (see Section \ref{section:source selection}) using the $\text{F606W} - \text{F814W}$ color, which brackets the $4000\ \text{\AA}$ break at the cluster redshift.

Carried by a spaceborne telescope, the ACS CCDs have been bombarded by high-energy particles. The particles have damaged the detectors and created traps resulting in charge transfer inefficiency (CTI). The CTI is prevalent and without correction would severely hamper our study. Using the latest pixel-based method, the STScI pipeline corrects for CTI in the ACS filters \citep{ubeda2012}. This automatic correction has been shown to be sufficient for weak-lensing analysis in \citet{jee2014weighing}.
For each of the \textit{HST}/ACS filters, the frames are distortion corrected, cleaned of cosmic rays and stacked using MultiDrizzle \citep{koekemoer2003}. The estimated alignment error of the stacked images is $\mytilde1$\% of a pixel, which is below the threshold required in galaxy cluster weak-lensing application. The images are ”drizzled” using a Lanczos3 kernel to a final pixel scale of $0\farcs05$ pixel$^{-1}$. For additional details of \textit{HST} data reduction, we refer readers to our previous papers \citep[e.g.,][]{jee2014weighing, jee2016toothbrush, golovich2017}.

\subsection{Keck/DEIMOS Data Reduction} \label{section:spectroscopy}
\begin{figure}[!htb]
\centering
\includegraphics[width=0.4\textwidth]{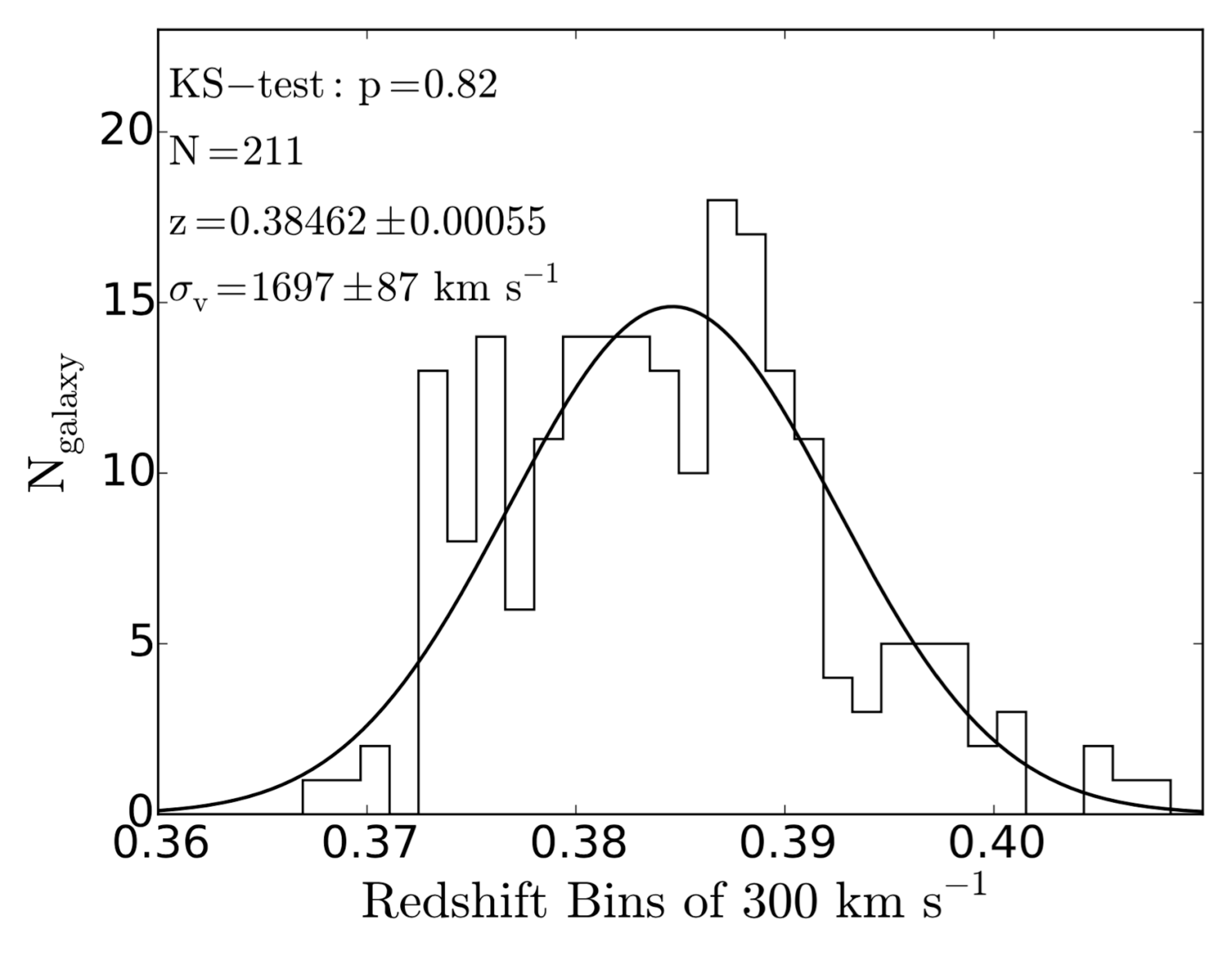}
\caption{Redshift distribution of 211 spectroscopically confirmed galaxies between redshift 0.36 and 0.41. A KS-test shows that the galaxies are well represented by a Gaussian distribution with no significant line-of-sight difference in substructures. The mean redshift of the cluster is found to be $0.385$ and the velocity dispersion $\sigma_\text{v}=1697\pm87\ \text{km}\ \text{s}^{-1}$.}
\label{fig:kstest}
\end{figure}

Spectroscopic observations (PI: W. Dawson) were done with the DEep Imaging Multi-Object Spectrograph (DEIMOS) mounted on the Keck II Telescope. Here we will highlight the spectroscopic procedure and refer the reader to \cite{dawson2015} and Golovich et al. (in prep.) for a thorough description. The data were collected with three slit masks on 2015 February 16 using the Subaru SuprimeCam photometric images for target selection. A 1,200 line mm$^{-1}$ grating was used with $1\arcsec$ wide slits, which resulted in a pixel scale of $0.33\ \text{\AA}\ \text{pixel}^{-1}$ and a resolution of $\mytilde 1\ \text{\AA}$. From this, we obtained secure spectroscopic redshifts for 211 cluster member galaxies in a redshift range of $0.36 < z < 0.41$. Figure \ref{fig:kstest} is the redshift distribution of the cluster members. The mean redshift of the cluster member galaxies is $z=0.385$ with a velocity dispersion of $\sigma_v = 1697\pm87\ \text{km}\ \text{s}^{-1}$. A Kolmogorov-Smirnov (KS) test compares the cumulative distribution of the galaxy redshifts with that of a Gaussian function and the result, $p=0.82$, suggests that the distribution is consistent with a Gaussian. This may suggest that the substructures of the cluster posses negligible line-of-sight velocity differences. We found no additional clustering in the range $0.15 < z < 0.90$.

\subsection{Object Detection and Photometry}
The main noise contribution to weak-lensing analyses is the large intrinsic ellipticity dispersion of galaxies. Therefore, one would like to have as many galaxies in the shape catalog as possible, including faint and small galaxies, as long as we can control systematic errors (e.g., noise bias). To achieve this, we do photometry with \texttt{SExtractor} in dual-image mode. The merit of dual-image mode is that it uses one image for detection and a second for photometry, allowing photometry to be performed consistently with the same isophotal aperture for each object across different filters. For the Subaru data, we utilize the deeper, $r$-band mosaic as the detection image and perform photometry on both the $g$- and $r$-band mosaics. For \textit{HST} data, a detection image is created by combining the F435W, F475W, F606W, and F814W mosaic images and photometry is done on F606W and F814W. A band specific weight image created by \texttt{SExtractor} during photometric data reduction is provided for detection and an rms image is provided for each photometry run. The rms image is created by multiplying the band specific weight image by the background rms of its mosaic image and masking spurious pixels. For object detection, we require objects to have a pixel value greater than two times the sky rms and to have at least five connected pixels. The settings for deblending are optimized to detect weakly-lensed galaxies. \texttt{DEBLEND\_NTHRESH} is set to 32 and \texttt{DEBLEND\_MINCOUNT} to $10^{-4}$ to maximize detection of overlapping objects.

\begin{figure}[!htb]
    \centering
    \includegraphics[width=0.45\textwidth]{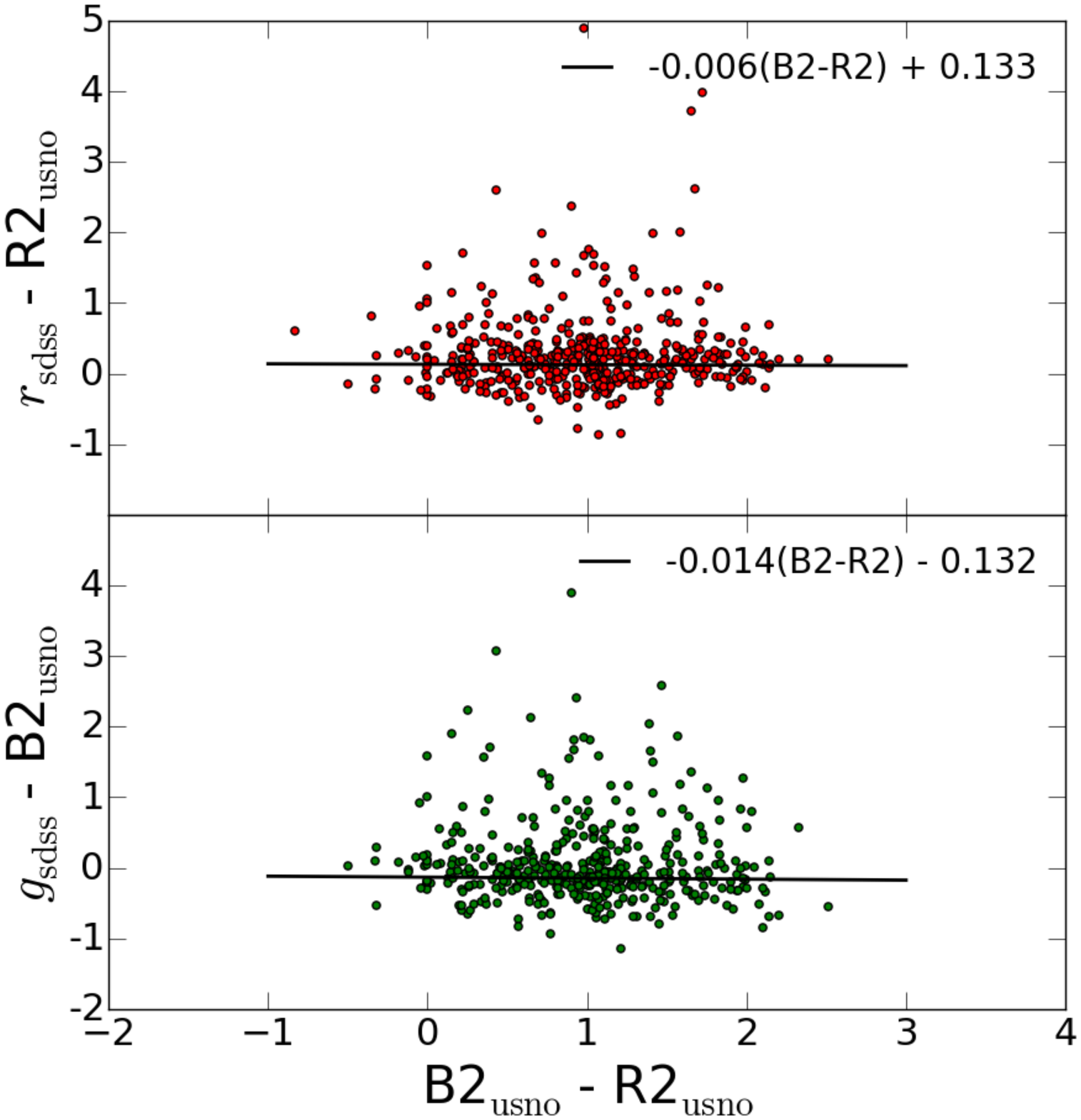}
    \caption{\label{fig:sdss_usno} Color-color diagrams used to calibrate the SDSS filters to the USNO reference catalog. Stars are matched in a common field between the SDSS and USNO catalogs and a linear fit to the color-color relation is found. The linear relation is applied to the stars in the USNO catalog that are matched to the field of PLCK~G287.0+32.9 in order to find the magnitude zeropoint. $3\sigma$ clipping is applied to remove outliers. \emph{Top:} Transformation between SDSS $r$ and USNO R2 filters. \textit{Bottom:} Transformation between SDSS $g$ and USNO B2 filters.}
\end{figure}

The Subaru photometric zero-point is calibrated by matching detected objects to an external star catalog. As our SuprimeCam photometric data are observed through the $g$- and $r$-band filters, which have similar throughputs to those of the SDSS filters, it would be ideal to directly use SDSS data to calibrate \texttt{SExtractor} magnitudes. However, PLCK~G287.0+32.9 is outside the SDSS sky coverage. Thus, we use both an SDSS catalog and a USNO-B1 catalog. First, we choose a field that has overlapping regions in the SDSS and USNO-B1 surveys. We then match stars in the field and plot a color-color diagram (Figure \ref{fig:sdss_usno}) to find a relation between the $g_{\mathrm{sdss}}$ and $r_{\mathrm{sdss}}$ filters and the B2$_{\mathrm{usno}}$ and R2$_{\mathrm{usno}}$ filters. The transformations are:
\begin{equation}
\begin{split}
r_{\mathrm{sdss}} &= -0.006(B2_{\mathrm{usno}}-R2_{\mathrm{usno}}) + 0.133 + R2_{\mathrm{usno}}\ ,\\
g_{\mathrm{sdss}} &= -0.014(B2_{\mathrm{usno}}-R2_{\mathrm{usno}}) - 0.132 + B2_{\mathrm{usno}}.
\end{split}
\end{equation}
These transformations allow us to find the SDSS equivalent magnitudes of the stars in the USNO catalog, which are then used to calibrate the SuprimeCam instrumental magnitudes. We also correct for Milky Way dust extinction with estimates provided by \cite{schlafly2011}.

The \textit{HST} zero-point calibration is more straightforward. We determine the AB zero-point using the header values for \texttt{PHOTPLAM} and \texttt{PHOTFLAM} and a dust correction is applied from \cite{schlafly2011}. 

\section{Point Spread Function Modeling} \label{section:psf modeling}
A critical step to recovering the weak-lensing signal is to model and remove the effect of the point spread function (PSF). In this analysis, we utilize the principal component analysis (PCA) method to model the PSF. As \cite{2007jeePCA} showed, the PCA approach accounts for small and large scale structures of the PSF better than the wavelet and shapelet methods. Additional merits are that PCA derives the basis functions from the data set itself and requires fewer components than the other two methods to model the PSF. Here we describe the Subaru $r$-band PSF modeling method in detail and follow it with a few comments on the alterations used for the \textit{HST} F814W PSF modeling.  

\begin{figure}[!htb]
    \centering
    \includegraphics[width=0.45\textwidth]{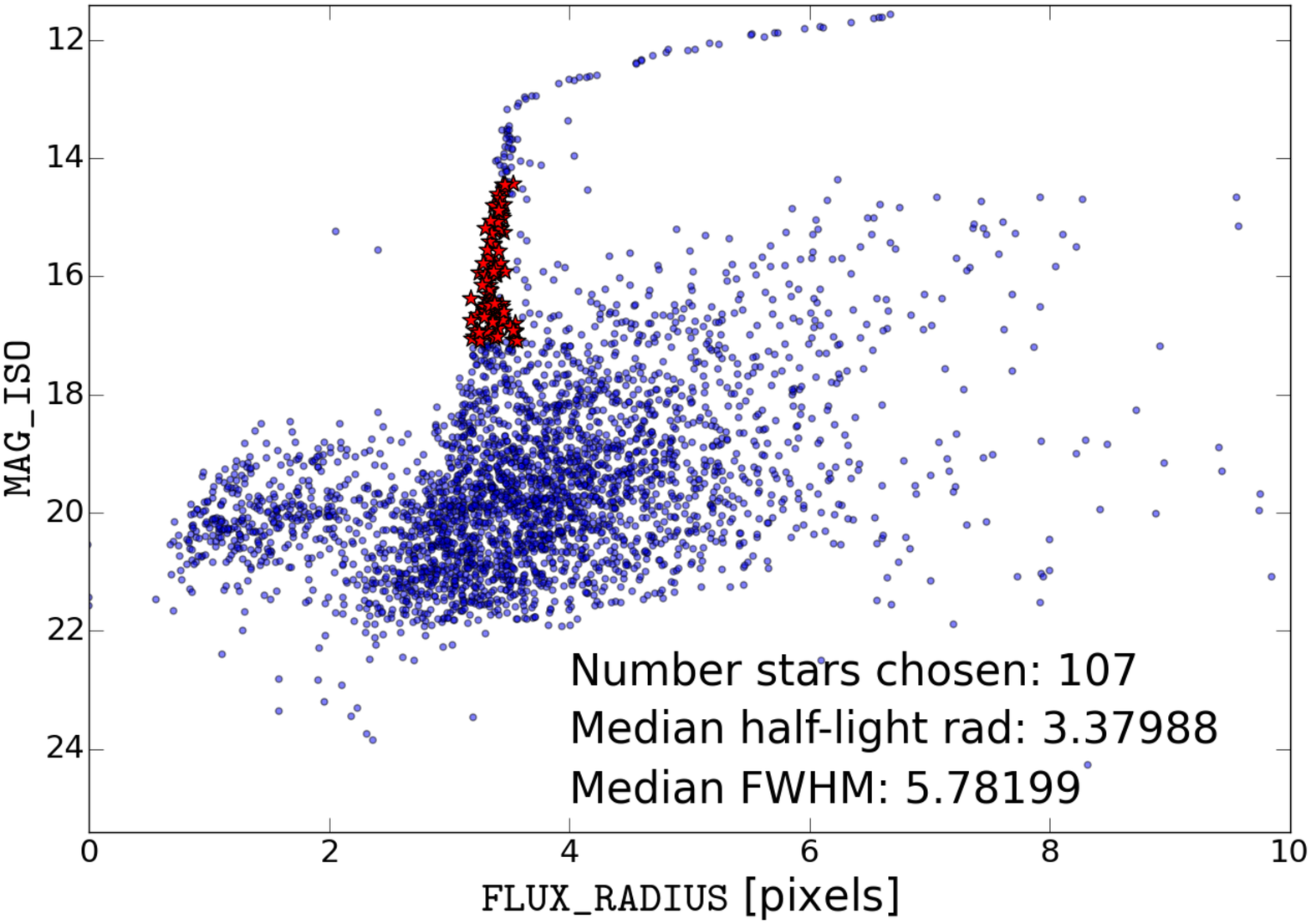}
    \caption{Star selection example. Stars used for PSF characterization are selected based on their magnitude and half-light radius as measured by \texttt{Source Extractor}. Saturated stars (non-linear response region) are avoided as their image does not represent the PSF. The stars selected for PSF measurement are marked with red stars and clearly follow the stellar locus.}
    \label{fig:star_select}
\end{figure}

The Subaru PSF varies from frame to frame as well as within each frame and the variation of the PSF is inherited by the co-added mosaic image from each of its component frames. Therefore, the PSF must be modeled for each frame individually and accumulated to a final PSF to be used for galaxies in the mosaic image. 

\begin{figure*}[!htb]
    \centering
    \includegraphics[width=\textwidth]{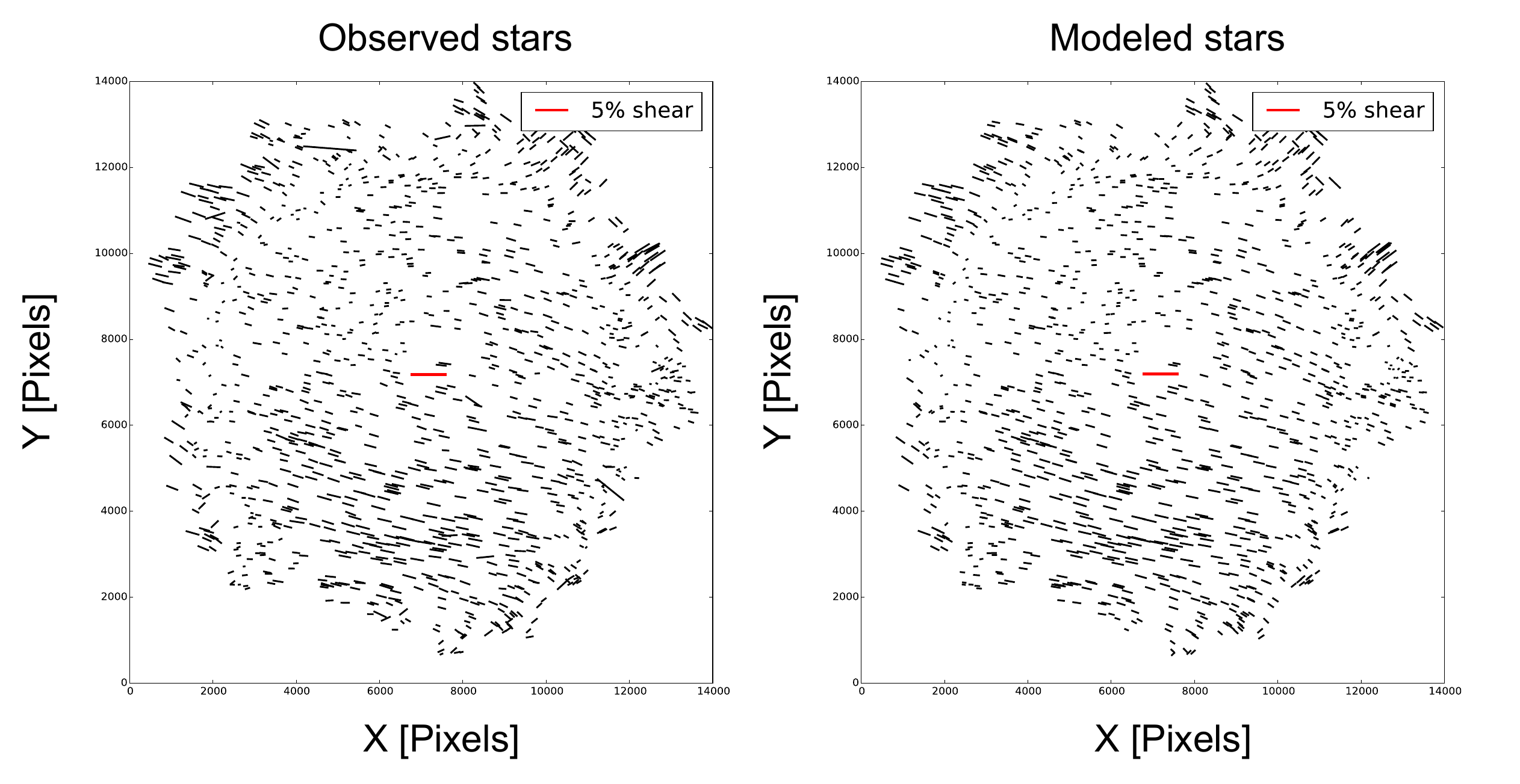}
    \caption{\label{fig:star_ellip} \emph{Left:} Spatial variation of the measured ellipticities of stars that are selected for PSF modeling from Subaru observations. The orientation and length of the ``whiskers'' represent the direction and magnitude of shape distortion by the atmosphere and telescope. The Subaru PSF varies across the field of view as well as between component frames. \emph{Right:} Modeled spatial variation of the Subaru PSF. The PSF is modeled for each component frame of the mosaic image and accumulated into a final PSF. The red line represents a 5\% shear from circular. Comparison of the two figures shows the agreement between the measured stellar ellipticity and the corresponding PSF model for stars.}
\end{figure*}

The first step in our PSF modeling is to select a robust sample of stars to measure the PSF from. Stars are identified in the RESAMP frames that are created during the \texttt{SWarp} step of our data reduction. The star sample is constrained by half-light radius, magnitude, maximum flux, and ellipticity ($e=(a-b)/(a+b)$, where $a$ and $b$ are the semi-major and -minor axes, respectively). Figure \ref{fig:star_select} shows the magnitude as a function of flux radius for objects in one of the RESAMP frames. In this particular frame, the stellar locus resides around 3.4 pixels and trails off where the CCD becomes non-linear with saturation. This should not be confused with the brighter-fatter effect (for recent work see \citealt{lage2017brighterfatter}), which is seen in the figure as a gradual increase in the flux radius of stars with decreasing magnitude. The stars used to model the PSF are marked with red stars. Typically, each frame provides approximately 100 ``good'' stars. It is assumed that the stars are point sources (i.e., delta function) and any shape is due to convolution with the PSF. To observe the shape of stars, postage stamp images (21 $\times$ 21 pixels$^2$) of each star are cut from the mosaic and fit with a Gaussian weighted quadrupole. The direction and magnitude of the ellipticity of the stars is shown in the left panel of Figure \ref{fig:star_ellip}. These ellipticities describe the variation of the PSF across the mosaic image. 

\subsection{Principal Component Analysis}
With the robust sample of stars determined, we can now model stellar PSFs and use them to construct galaxy PSFs. We characterize the PSF using PCA on the stars in each of the RESAMP frames. PCA is an orthogonal linear transformation of a data set to a new basis of vectors (principal components), ordered by their decreasing degree of variance. One method to implement an orthogonal linear transformation is PCA with a covariance matrix. Briefly, the steps of PCA applied to a single RESAMP frame are:

\begin{enumerate}
    \item Determine the mean stellar PSF and residual PSF
    \item Calculate the covariance matrix of the residual
    \item Compute eigenvectors and eigenvalues of the covariance matrix
    \item Select the eigenvectors that describe the majority of the PSF variation
    \item Project the stellar images onto the new orthogonal basis (principal components)
\end{enumerate}

The first step is to determine the mean stellar PSF and compute the residual PSF ($\text{star} - \text{mean}$) for each star. Stellar postage stamp images (21 pixels $\times$ 21 pixels) are cut from the RESAMP image. RESAMP images are ideal for measuring the PSF because they have been rotated, shifted, and distortion corrected as needed for the co-added mosaic. Next, the stellar postage stamps are stacked into a data cube of $N$ (stars) $\times$ 21 (pixels) $\times$ 21 (pixels). When creating the data cube, sub-pixel shifts are applied to each postage stamp to ensure the peak of the stellar intensity is located at the center pixel. As explained in \cite{2007jeePCA}, the variance of the PSF would be dominated by the scattered peak positions if centering was not applied. Bicubic interpolation is used to achieve the sub-pixel shifts because it creates fewer interpolation artifacts than higher order windowed sinc interpolation schemes \citep{2007jeePCA}. Once centered, the flux of each star is normalized. At this point, we remove a dimension from the data cube by flattening each stellar image into a 441 component one-dimensional array of pixels, $M$. This results in a matrix that is $M$ columns (pixels) $\times$ $N$ rows (stars). We calculate the mean PSF by averaging along each column and then subtract the mean from each row to get the residuals, $R$. These residuals represent the deviation of each stellar PSF from the mean PSF for a single RESAMP frame. 

Next, the covariance matrix for $R$, the residuals, is calculated and the eigenvalues and their corresponding eigenvectors are determined from the covariance matrix. The eigenvalues are the variance of the residual and the eigenvectors are the principal components. As shown in \cite{2007jeePCA}, the majority of the variance exists in the 21 components with the highest eigenvalues. This allows us to describe the residual in 21 components without sacrificing precision. The last step of the PCA is to project the residuals $R$ onto the orthogonal basis formed by the 21 principal components.

\subsection{Modeling the PSF}
The objective of modeling the PSF is to create images that represent the PSF of galaxies in the stacked mosaic image, where the weak-lensing signal is measured. Therefore, the variation of the PSF as measured from stars needs to be applicable to galaxies. We achieve this by fitting the spatial variation of each of the 21 components of the projected $R$ with a 3rd order polynomial and storing the 10 coefficients, $d_n$. The coefficients represent the variation of the PSF from the mean PSF across an individual RESAMP frame. The PCA and polynomial fitting process is repeated for each RESAMP frame that was co-added to create the mosaic. The PSF at any location (x, y) of the mosaic is constructed for each RESAMP frame and then stacked to a final PSF. The reconstruction of the $k^{th}$ object PSF, before stacking, has pixel values:

\begin{equation}
    C_k(i,j) = \sum^{20}_{n=0} a_{kn}P_n(i,j) + T(i,j),
\end{equation}
where $a_{kn}$ are the coefficients of the $k^{th}$ object derived from the $d_n$ coefficients as
\begin{equation}
a_{kn} = \sum^{l+m\leq 3}_{l,m=0} d_{nlm}x^ly^m,
\label{polyfit}
\end{equation}
with $x$ and $y$ the mosaic image coordinates of the object. The polynomial fit is found by projecting the coefficients onto the principal component basis, $P_n(i,j)$, and the model PSF is constructed by adding the mean PSF, $T(i,j)$. The $k^{th}$ object PSF from each model is saved to a fits file to be used in shape fitting. The right panel of Figure \ref{fig:star_ellip} shows the quadrupole measured ellipticity of the modeled stellar PSFs. Comparing the panels of the figure, the ellipticity of modeled stellar PSFs is in good agreement with the observed ellipticity.

Modeling of the PSF for the \textit{HST} frames is also done using a PCA approach. However, unlike the Subaru observations, the \textit{HST} frames do not have the necessary number of high S/N stars to extract the PSF information from the science frames. Thus, we match PSFs for each frame from a PSF library created using external stellar field images \citep{2007jeePCA}. As \cite{2007jeePCA} showed, this is possible because the variation in the \textit{HST} PSFs are primarily caused by the focus of the telescope. The PSF variation pattern becomes nearly identical when two observations are taken under the same focus offset.

\section{Weak Gravitational Lensing Theory} \label{section:lensing theory}
Since the first detection by \cite{tyson1990detection}, weak gravitational lensing has become a powerful tool to investigate the mass of galaxy clusters. Immense progress has been made in understanding the systematic uncertainties that are introduced by the telescope and atmosphere. Along with these advances, the nature of weak lensing has cemented it as a distinct tool in galaxy cluster mass determination. In this section, we will give an overview of weak lensing theory.  For a more detailed description of weak lensing theory we refer the reader to reviews by \cite{bartelmann2001weak} or \cite{2013hoekstra}.

Weak lensing by a galaxy cluster occurs when light from galaxies beyond the cluster (source galaxies) is deflected as it passes through the cluster gravitational potential. The simplest way to model gravitational lensing is to consider the galaxy cluster as a thin lens. This approximation holds because the size of a galaxy cluster is small compared to the distance from observer to cluster. In the thin lens approximation, the gravitational potential is analogous to the index of refraction for media. As light passes through the galaxy cluster, it follows Fermat's principle, following the path of extrema in travel time. This introduces a shift in the apparent location of the source galaxy on the sky. Considering the geometry of the system, we can formulate the relation between the true source position, $\boldsymbol{\beta}$, and the observed position, $\mathbf{x}$, as

\begin{equation}
\boldsymbol{\beta} = \mathbf{x} - \boldsymbol{\alpha}(\mathbf{x}).
\end{equation}
The scaled deflection angle, 
\begin{equation}
\boldsymbol{\alpha}(\mathbf{x}) = \frac{1}{\pi} \int \kappa(\mathbf{x'})\frac{\mathbf{x}-\mathbf{x'}}{|\mathbf{x}-\mathbf{x'}|^2}d^2\mathbf{x},
\end{equation}
is the gradient of the deflection potential, $\pmb{\alpha} = \nabla \psi$, where

\begin{equation}
\psi(\mathbf{x}) = \frac{1}{\pi} \int \kappa (\mathbf{x'})\ln|\mathbf{x} - \mathbf{x'}|d^2\mathbf{x'},
\end{equation}
and the convergence $\kappa$ (defined below), is the dimensionless surface mass density. Through a Green's function, it can be shown that the deflection potential satisfies the Poisson equation $\nabla^2\psi(\mathbf{x})=2\kappa(\mathbf{x})$. 

The convergence ($\kappa$) is one of two lens properties used to quantify the gravitational lensing effect. It causes an isotropic focusing of the light rays, which magnifies the image. It is defined as the projected surface mass density ($\Sigma$) of the cluster normalized by the critical surface mass density ($\Sigma_c$) at the cluster redshift:

\begin{equation}
\kappa = \frac{\Sigma(\mathbf{x})}{\Sigma_c}, \ \Sigma_c = \frac{c^2 D_s}{4\pi G D_l D_{ls} } ,
\label{eq:kappa}
\end{equation}
where $G$ is the gravitational constant, $D_l$ is the angular diameter distance to the lens, $D_s$ is the angular diameter distance to the source, $D_{ls}$ is the angular diameter distance from lens to source, and $c$ is the speed of light. The second lens property is shear ($\gamma$) and is caused by the tidal gravitational field. The shear anisotropically distorts the shape of galaxy images. In the weak-lensing regime ($\kappa \ll 1$, $\gamma \ll 1$), gravitational lensing results in small distortions and single galaxy images. 

For source galaxies of angular size much smaller than the variation of lens properties the coordinate transformation by weak gravitational lensing may be expressed by the Jacobian matrix $A$:

\begin{equation}
A = (1-\kappa)\begin{bmatrix}
1-g_1 & -g_2 \\ -g_2 & 1 + g_1
\end{bmatrix} ,
\label{eq:jacobian}
\end{equation}
where
\begin{equation}
    g=\frac{\gamma}{1-\kappa}
    \label{eq:reduced_shear}
\end{equation}
is the reduced shear. Joining the two components of the reduced shear gives a complex representation of the shear, $g = g_1 + ig_2$. Positive and negative values of $g_1$ distort the shape of the galaxy along the $x$-axis and $y$-axis directions, respectively, whereas, positive and negative values of $g_2$ distort the shape along the $y=x$ and $y=-x$ directions, respectively. The effect of the transformations change a circle into an ellipse. As shown in \cite{kaiser1993mapping}, the shear and convergence are related by the following convolutions:

\begin{equation}
\kappa(\mathbf{x}) = \frac{1}{\pi} \int D^*(\mathbf{x} - \mathbf{x'})\gamma(\mathbf{x'})d^2\mathbf{x'} ,
\label{eq:kappa_convolve}
\end{equation}

\begin{equation}
\gamma(\mathbf{x}) = \frac{1}{\pi} \int D(\mathbf{x} - \mathbf{x'})\kappa(\mathbf{x'})d^2\mathbf{x'} ,
\end{equation}
where $D(\mathbf{x}) = -1/(x_1 - ix_2)^2$ is the convolution kernel. In the weak-lensing regime ($\kappa \ll 1$), the reduced shear is approximately equal to the shear, allowing a reconstruction of the surface mass density through measurement of the reduced shear. To measure the reduced shear, we do a statistical analysis of the observed shape of galaxies beyond the cluster. The ellipticity of a galaxy image is:
\begin{equation}
\epsilon \approx \epsilon_{\mathrm{intrinsic}} + g, 
\end{equation}
where $\epsilon_{\mathrm{intrinsic}}$ is the source galaxy intrinsic ellipticity. Assuming the source galaxies are randomly oriented, their complex ellipticity should sum to zero over a large statistical sample. It is then given that the averaged ellipticity of images is the reduced shear:
\begin{equation}
    \langle \epsilon\rangle\approx g .
    \label{eq:average_eq_redshear}
\end{equation} In the following section, we will discuss the steps required to get the reduced shear by measurement of galaxy shapes.  

\section{Shape measurement with PSF correction} \label{section:shape measurement}
In weak lensing, it is critical to accurately measure the shapes of the source galaxies so that the weak-lensing distortion can be quantified with minimal systematic errors. A popular method to measure shapes was proposed by \citet*[][KSB hereafter]{kaiser1995method}. The KSB method and its many variations use the quadrupole moments of galaxies and PSFs to estimate the shapes of galaxies. The limitations of this method, such as the uncertainty arising from photon noise and dense fields, have been outlined in \cite{kaiser2000ksb}. 

Another method to measure galaxy shapes is to fit their light profile with an analytic model. As highlighted in the GREAT3 challenge \citep{Mandelbaum2015great3}, fitting a galaxy light profile with an inexact model introduces ``model bias'' into the measured shear. We choose to model galaxy shapes by fitting the light distribution with an elliptical Gaussian function. To account for model bias, as well as noise bias, we apply a multiplicative correction factor of 1.15 to calibrate our ellipticities. The correction factor is derived from simulations using real galaxy images as described in \citet{jee2013cosmic}. The elliptical Gaussian function is 

\begin{equation}
\begin{split}
G(x, y) = A_0 + A_1 \exp & \left[-\frac{(\Delta x\cos\theta-\Delta y\sin\theta)^2}{2\sigma_x^2}\right.\\
    & \left. + \frac{(\Delta x\sin\theta+\Delta y\cos\theta)^2}{2\sigma_y^2}\right],
\end{split}
\label{eq:gauss}
\end{equation}
where $\Delta x$ and $\Delta y$ are $x - x_0$ and $y - y_0$. The parameters of the Gaussian function are background flux ($A_0$), peak flux ($A_1$), centroid ($x_0, y_0$), variance ($\sigma_x^2, \sigma_y^2$), and orientation angle ($\theta$). It is well known that the elliptical Gaussian function is a rough approximation of the galaxy light profile. However, weak-lensing is a statistical study of minute galaxy shape distortions and the most effective means to achieving high S/N is to include as many galaxies as possible. For this reason, we choose to fit galaxies with a model that requires few free parameters. \cite{jee2013cosmic} explain that a S\'{e}rsic profile (a better model for galaxy light distribution) increases the uncertainty of the measurement by including more pixels with low signal and by having more free-parameters to fit. Therefore, although using the Sersic profile decreases model bias, the larger uncertainty due to additional parameters, in turn, increases noise bias. 

Each object in the mosaic image is fit with its own elliptical Gaussian function. To do so, a square postage stamp image of each object is created from the mosaic. The dimensions of the postage stamp are chosen to be 4 x \texttt{A\_IMAGE} per side, where \texttt{A\_IMAGE} is the semi-major axis determined by \texttt{SExtractor}. This provides enough room to fit the galaxy profile while at the same time limiting contamination from other sources.  

Before fitting the model to each galaxy, it is critical to consider the PSF. We use forward modeling to account for the PSF. This is done by convolving the model Gaussian function (Equation \ref{eq:gauss}) with the model PSF prior to fitting the object. The convolved model is:

\begin{equation}\label{eq:conv}
M(x,y) = G(x,y) \otimes C(x,y),
\end{equation}
where $G$ is the Gaussian model and $C$ is the PSF model. An iterative minimization method is applied to find the best fit elliptical Gaussian model for the target object in each postage stamp. For this step, we use \texttt{MPFIT} \citep{markwardt2009}, which performs least squares minimization using a Levenberg-Marquardt method. \texttt{MPFIT} accepts initial conditions and allows constraints for all parameters of the function being fit. We fix the background and centroid to the \texttt{SExtractor} measured values in order to minimize the free parameters and do not impose any additional constraints on the fit. The algorithm minimizes the difference between the PSF convolved model $M_i(x,y)$ and the observed image $O_i(x,y)$:

\begin{equation}
\chi^2 = \sum_{i=1}^{n} \frac{(O_i(x,y) - M_i(x,y))^2}{\sigma_i^2}.
\end{equation}
For the variance of the $\chi^2$, we provide an rms postage image. Segmentation and rms postage images are cut from the rms and segmentation mosaics output from \texttt{SExtractor}. These images provide a means to disregard pixels from the postage image that do not belong to the current target object, as defined by \texttt{SExtractor}. This is accomplished by setting the rms value of pixels selected from the segmentation map that do not belong to the current target object to an arbitrarily large number ($10^6$). This effectively hides their flux from the fitting algorithm. 

The outputs from the $\chi^2$ fit are the seven elliptical Gaussian parameters, status of fit, reduced $\chi^2$, and the $\chi^2$ error. The status is a notification of the success of fit and the $\chi^2$ error is the 1$\sigma$ error for each parameter with an assumed Gaussian prior. From the outputs, we determine the semi-major axis ($a$), semi-minor axis ($b$), and the orientation angle ($\theta$, measured counter-clockwise from the x-axis to the semi-major axis) for each object. The complex ellipticity can then be defined as $e = e_1 + ie_2$ with
\begin{equation}
\begin{split}
    e_1 = \frac{a-b}{a+b}\cos(2\theta), \\
    e_2 = \frac{a-b}{a+b}\sin(2\theta).
\end{split}
\end{equation}
The ellipticity error is found by propagating the errors that are output for each of the parameters. Finally, a catalog of the ellipticities for all objects detected from the mosaic image is accumulated. 
Residual multiplicative biases in shears derived from these shape estimates could be present because the simulations we used for calibration are not perfect representations of our data. Nevertheless, such biases are likely small (at the level of a few per cent) and thus subdominant to the statistical uncertainty of this study.

\section{Source selection} \label{section:source selection}
The geometry of gravitational lensing requires that the source galaxies be located at a greater distance than the cluster in order for their light to be distorted by the galaxy cluster potential. It is thus expected that these galaxies, on average, appear fainter than the cluster member galaxies. Also, due to the hierarchical formation of the Universe, we expect the average cluster galaxies to be more evolved and thus redder. These two characteristics are the basis of a color-magnitude diagram and provide the primary selection criteria for our background source catalog. 

\begin{figure}[!htb]
    \centering
    \includegraphics[width=0.5\textwidth]{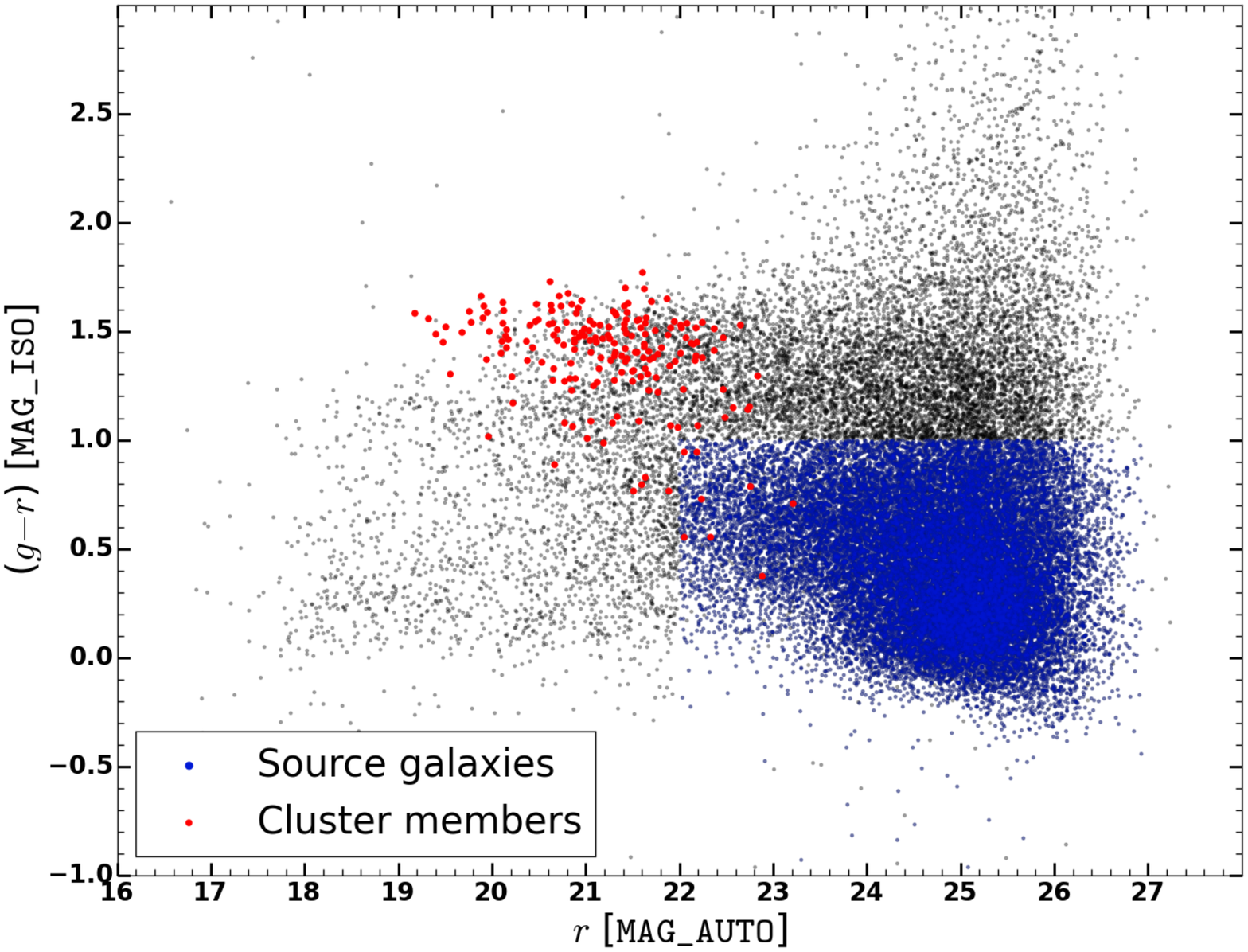}
    \caption{Color-magnitude relation for objects detected in the PLCK~G287.0+32.9 field from Subaru photometry. All magnitudes have been extinction-corrected. The red sequence is selected using the 4000 \AA\ break, which is bracketed by the $g$- and $r$-band filters at $z=0.385$. Cluster members confirmed by KECK/DEIMOS spectroscopy are highlighted red and form a red sequence at $g-r\mytilde1.5$. Galaxies that are included in the weak-lensing source catalog (blue) are chosen to span a region fainter and bluer than the red sequence. Spectroscopically confirmed cluster members are removed from the source catalog.}
    \label{fig:CMD}
\end{figure}

Figure \ref{fig:CMD} is the color-magnitude diagram for Subaru observations of PLCK~G287.0+32.9. The horizontal axis is the $r$-band \texttt{MAG\_AUTO} and the vertical axis is $g-r$ \texttt{MAG\_ISO} color. The red markers indicate cluster members confirmed by DEIMOS spectroscopy. Near $g-r \sim 1.5$ the confirmed cluster members form a red sequence. Combining the color-magnitude relation and the spectroscopically confirmed cluster members, we create a Subaru cluster member catalog. The catalog contains sources whose $g-r$ color is between 1.2 and 1.7 and $r$ magnitude between 19 and 23. From this catalog, we remove spectroscopically confirmed non-cluster members. This cluster member catalog is used to create cluster luminosity and number density maps. 

Next, we select the our source galaxies by applying the following constraints:

\begin{itemize}
    \item $22 < r < 27$
    \item $-1 < g - r < 1$
    \item semi-minor axis, $b > 0.2$ pixels
    \item ellipticity, $e < 0.9$
    \item ellipticity error $< 0.3$.

\end{itemize}
The source galaxies selected in this way are fainter and bluer than the red sequence. To separate stars from background galaxies, the semi-minor axis after deconvolution is required to be larger than 0.2 pixels. This excludes objects that have been improperly fit with an extremely elongated ellipse or whose signal covers too few pixels to be from a sheared galaxy. Further star-galaxy separation is achieved by requiring the pre-seeing ellipticity to be less than 0.9 and the ellipticity error to be less than 0.3. The source galaxies are plotted with blue markers in the color-magnitude diagram. The total number of source galaxies is 27,089 galaxies ($\mytilde$27 galaxies arcmin$^{-2}$).

An \textit{HST} source catalog is created using the same constraints as the Subaru catalog with the following adjustments to the magnitude and color cuts:

\begin{itemize}
\item $22 < F814W < 28$
\item $-0.5 < F606W - F814W < 0.8$.
\end{itemize}
The mean number density of sources in the \textit{HST} source catalog is $\mytilde 110\ \text{arcmin}^{-2}$. An \textit{HST} cluster member catalog is created by selecting galaxies with magnitude ranging from 16 to 23 and $F606W-F814W$ color between 0.8 and 1.3. These constraints encapsulate the spectroscopically confirmed cluster members. 

Obviously, our source selection above does not remove blue cluster members. If significant, the contamination will dilute our lensing signal and thus lead to underestimation of the cluster mass. This issue has been discussed by a number of authors \citep{broadhurst2005, okabe2010, applegate2014, medezinski2017, melchior2017}. However, based on the number density excess test with respect to control fields, our previous analyses have shown that the dilution, if any, is much below the statistical errors. For example, Figure 4 of \cite{jee2014weighing} demonstrates that the source number density distribution in the ``El Gordo'' cluster at z = 0.87, obtained by removing only the red cluster members, is in good agreement with the statistics in GOODS-N and -S fields, which implies that the blue member contamination is negligible. More recently, a similar conclusion is made by \cite{jee2017} for two clusters at even higher redshift ($z \gtrsim1.5$). Because PLCK~G287.0+32.9 is at much lower redshift (z = 0.385) than these clusters, it is unlikely that the contamination becomes a critical issue in the current study.
\section{Redshift Estimation}

\begin{figure}[!htb]
    \centering
    \includegraphics[width=0.4\textwidth]{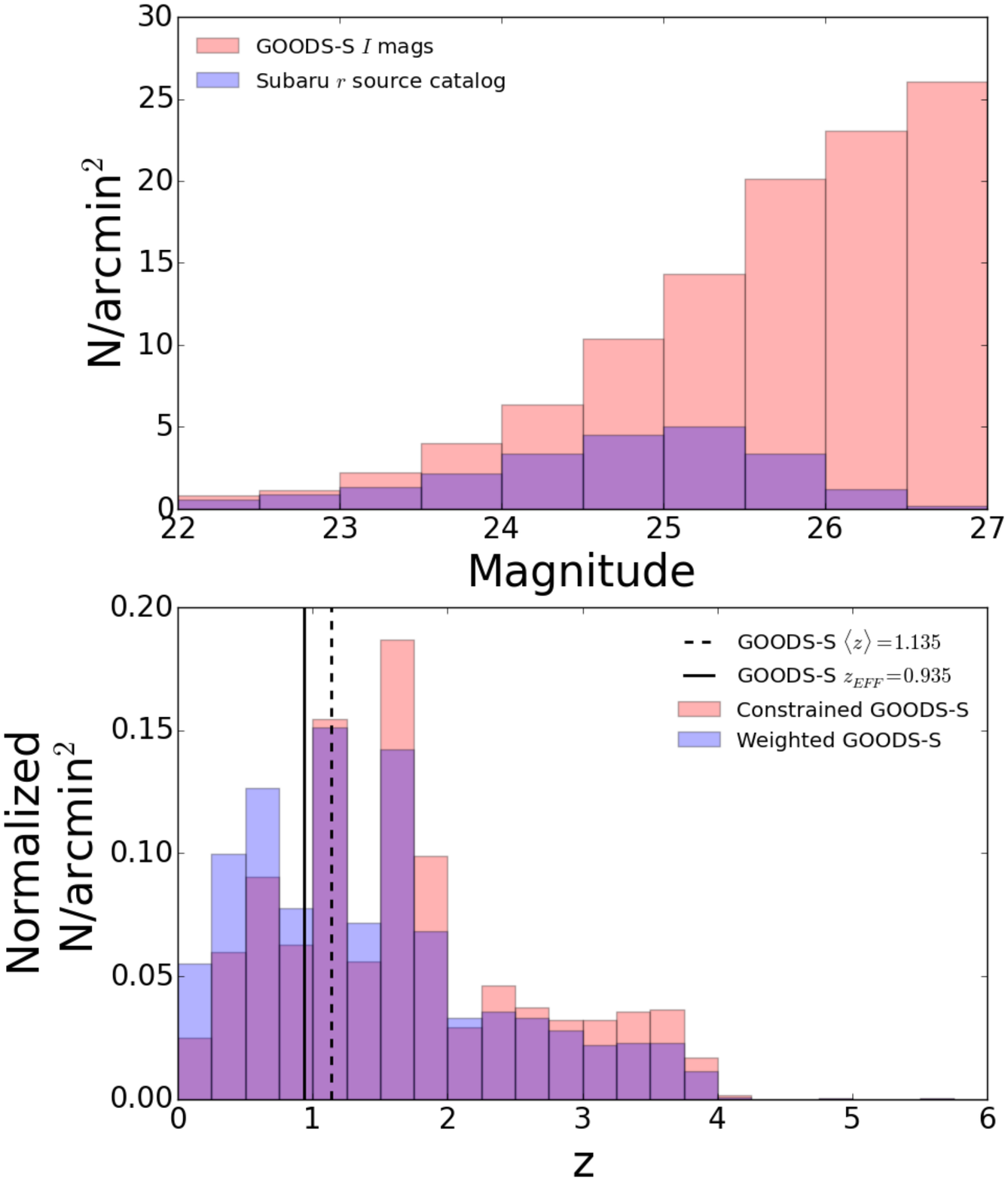}
    \caption{Estimation of the effective redshift for Subaru source catalog galaxies using the GOODS-S photometric redshift catalog of \cite{dahlen2010}. The GOODS-S catalog is modeled to represent the redshift distribution of our Subaru source catalog. \textit{Top}: Number density of objects in the color and magnitude constrained GOODS-S catalog (red bars) and the Subaru source catalog (blue bars) highlights the difference in depth between the two catalogs. \textit{Bottom}: Redshift distribution of the GOODS-S catalog before (red) and after (blue) correcting for depth and applying Equation \ref{eq:beta}.}
    \label{fig:effective_z}
\end{figure}

As shown in Equation \ref{eq:kappa}, the convergence depends on source redshift (more exactly, sensitive to the ratio of $D_{ls}/D_s$). We cannot determine photometric redshifts for source galaxies for PLCK~G287.0+32.9 based on two filters. Therefore, we estimate the redshift distribution of the source galaxies by modeling an external photometric redshift catalog to match our source catalog. We use the Great Observatories Origins Deep Survey (GOODS) South photo-$z$ catalog compiled by \cite{dahlen2010}. The GOODS-S catalog makes use of 12 photometric bands to determine redshifts and has a total sky coverage of $\mytilde 153\ \text{arcmin}^2$. 

The red bars in the top panel of Figure \ref{fig:effective_z} show the magnitude distribution of galaxies in the GOODS-S catalog after applying the filter-corrected constraints (Section \ref{section:source selection}) that were used to form the Subaru source catalog. The magnitude distribution of our Subaru source catalog galaxies (blue bars) shows a large departure from that of the constrained GOODS-S catalog at faint magnitudes because the GOODS-S imaging is much deeper than our Subaru imaging. We weight the GOODS-S catalog magnitude distribution to match our source catalog magnitude distribution. The weighted distribution is shifted to lower redshift, as the blue bars in the bottom panel show. However, foreground galaxies, which are not lensed, do not contribute to the lensing signal. To account for dilution of the lensing signal (Equation \ref{eq:kappa}) by foreground galaxies, we enforce the following to the GOODS-S catalog:

\begin{equation}
 \beta = max\left[\frac{D_{ls}}{D_s},0\right].
 \label{eq:beta}
\end{equation}
Using this definition of lensing efficiency gives an effective redshift. The effective redshift of the weighted GOODS-S sample that has been modeled to represent our Subaru source catalog is $ z_{\text{eff}}=0.935$ and $\langle\beta\rangle=0.523$. Representation of the redshift distribution of galaxies by a single number leads to an overestimation of the shear \citep[e.g.,][]{hoekstra2000}. The first order correction applied to the reduced shear to take the width of the redshift distribution into consideration when fitting a shear model is \citep{seitz1997}:
\begin{equation}
g' = \left[1 + \left(\frac{\langle\beta^2\rangle}{\langle\beta\rangle^2}-1\right) \kappa \right] g,
\end{equation}
where the variance of the distribution for the Subaru source catalog is $ \langle \beta^2 \rangle = 0.33$ and $\kappa$ is the modeled (iterated) convergence. 

We also use the GOODS-S catalog to estimate the redshift of the \textit{HST} source catalog. After applying Equation \ref{eq:beta} and correcting for depth, our \textit{HST} source catalog is estimated to have $\langle \beta \rangle = 0.545$ with an effective redshift of $z_{\text{eff}} = 0.996$. The width of the distribution is found to be $\langle \beta^2 \rangle = 0.35$.
We note that these estimates of $\left < \beta \right >$ have an uncertainty, in particular due to the small area of the GOODS-S data. Comparison with detailed studies of these effects \citep[e.g.,][]{gruen2017,jee2017} indicates that this is below a relative uncertainty of 5 per cent and thus subdominant compared to the statistical uncertainty of our mass estimates.

\section{Results}
\subsection{Surface Mass Density} \label{section:mass density}

Merging galaxy clusters, such as the Bullet Cluster, have been shown to have multi-modal mass distributions. This is not surprising as the merging timescale is of the order of giga-years. PLCK~G287.0+32.9 shows telltale signs of a merger as it hosts two radio relics and a radio halo. Therefore, it may be expected to contain a multi-modal mass distribution. To investigate the mass distribution, we create mass maps in the following manner.

\begin{figure*}[!htb]
    \centering
    \includegraphics[width=\textwidth]{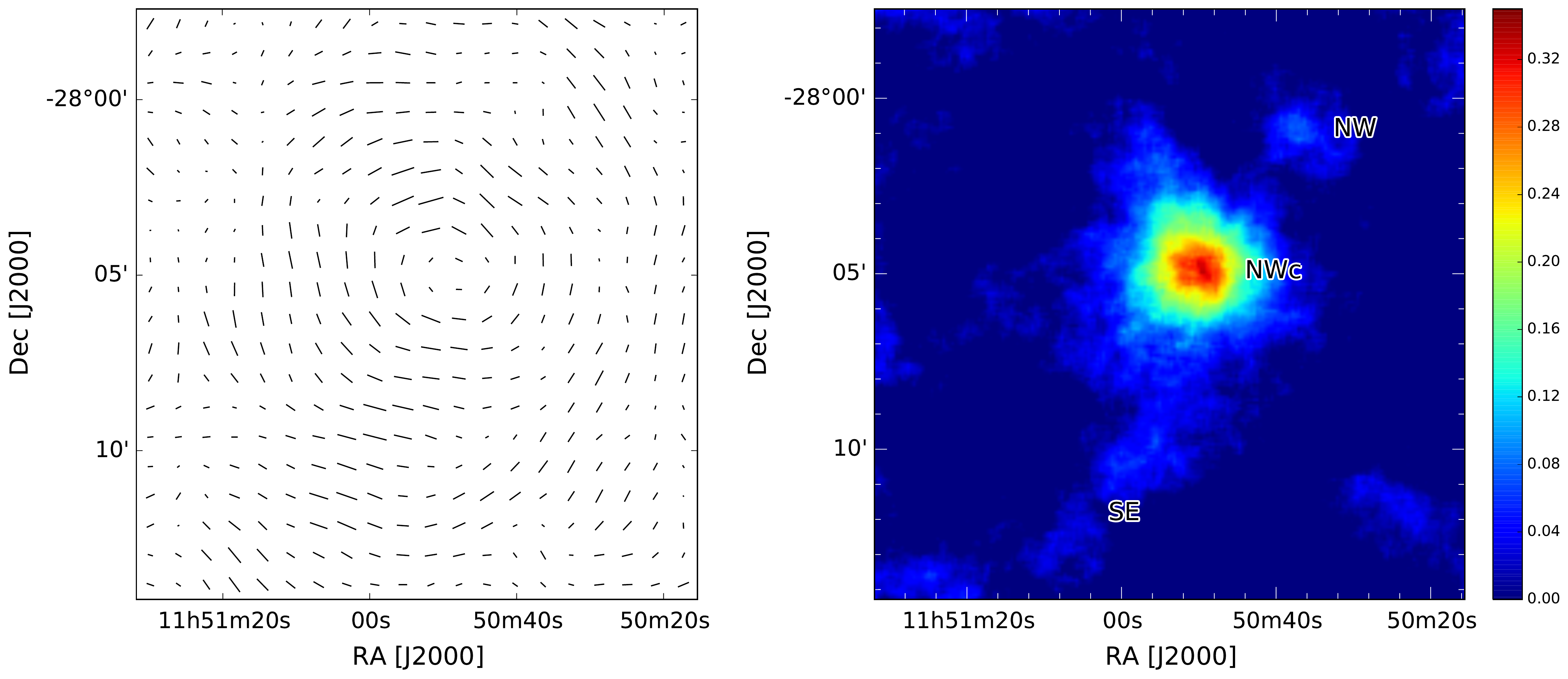}
    \caption{\label{fig:low_res} \textit{Left:} Whisker plot of the Subaru reduced shear determined by averaging the galaxy ellipticities within an $r=80\arcsec$ top-hat kernel. The direction and length of the whiskers represent the direction and magnitude of the reduced shear. The shear tends to be tangential to the cluster center and increase in magnitude towards the cluster center (around $11^\text{h}50^\text{m}45^\text{s}$, -28$^\circ05\arcmin$). \textit{Right:} Subaru convergence reconstructed by the inversion (Equation \ref{eq:kappa_convolve}) of the shear map. The mass reconstruction shows a clear peak. Low convergence detections are seen to the north-west and south-east.}
\end{figure*}

First, a shear map is created by spatially averaging the galaxy ellipticities from the source catalog. The left panel of Figure \ref{fig:low_res} is a low resolution visualization of the Subaru reduced shear created by averaging the source galaxy ellipticities within an $r=80\arcsec$ top-hat kernel. Each whisker represents the local magnitude and direction of the averaged ellipticity or, approximately, the reduced shear. This figure shows that the shear is predominantly oriented tangential to the cluster center (around $11^\text{h}50^\text{m}45^\text{s}$, -28$^\circ05\arcmin$) with the magnitude of the shear tending to decrease with distance from the center. A convergence map, as shown in the right panel of Figure \ref{fig:low_res}, is reconstructed by convolving the shear map with the transformation kernel (Equation \ref{eq:kappa_convolve}). As portrayed in Equation \ref{eq:kappa}, the convergence is the normalized projected surface mass density, which is dominated by dark matter. With this in mind, we will use the terms convergence, mass, dark matter, and projected surface mass density interchangeably.

The resolution of the mass map is sensitive to the size of the kernel used to average the source ellipticities. The primary concern when selecting a kernel size is to ensure the number of galaxies averaged over is sufficient enough that the relation $\langle e\rangle\ \approx\ g$ holds. At the same time, we wish to probe the details of the mass distribution to a precision that may allow the detection of substructure. With this in mind, we create two versions of convergence maps.

\subsubsection{Subaru Telescope Convergence Map}
As mentioned above, the Subaru convergence map is created with an $80\arcsec$ radius top-hat kernel. For each spatial bin, $\mytilde150$ source galaxies, on average, are used to find the reduced shear. The right panel of Figure \ref{fig:low_res} is the Subaru convergence reconstruction. It is important to note that the convergence values are subject to mass-sheet degeneracy \citep[see e.g.,][]{falco1985msd, schneider1994}, thus they should not be taken at face value but are useful to distinguish the distribution of mass. The Subaru convergence reconstruction shows a remarkably strong signal that dwarfs any nearby detections. This primary mass clump has a distinct global maximum (mapped in red color and labeled NWc). Around the peak, the convergence has a somewhat triangular shape with vertices pointing west, south-east, and north-east (mapped in green color). A lesser detection $\mytilde$2 Mpc to the north-west of the global maximum is a local maximum (labeled NW) that is found to have a statistical significance of $\mytilde 3\sigma$ (see below for bootstrap analysis). A second local maximum ($\mytilde3\sigma$), at approximately $11^\text{h}50^\text{m}55^\text{s}$, $-28^\circ10\arcmin$ (labeled SE), is found in the elongation to the south-east. Based on the convergence map, it can be inferred that NWc is the primary cluster and that the SE and NW convergence detections could be subclusters.

In the left panel of Figure \ref{fig:subaru_convergence}, the mass distribution is overlaid onto a Subaru color-composite image with color enhanced GMRT radio (green) and XMM X-ray (red) emissions. It is evident that NWc is within the X-ray emitting ICM. The NW mass clump is found to be further north-west than the NW radio relic. The SE convergence is peaked before the SE radio relic, which may be consistent with a scenario where a merging event between the primary cluster and this substructure created the SE radio relic. The convergence then stretches further south-east and crosses the western edge of the SE radio relic.

The right panel of Figure \ref{fig:subaru_convergence} compares the mass reconstruction with the $r$-band luminosity density distribution created using the Subaru cluster member catalog (Section \ref{section:source selection}). The luminosity density is smoothed with a FWHM $=50\arcsec$ Gaussian kernel. On the large scale, the luminosity distribution stretches in a north-west to south-east direction, similar to the convergence. The luminosity distribution shows that the primary luminosity peak is matched well with the NWc convergence map peak. This luminosity peak is dominated by the flux of the BCG. Trailing off to the south-east, a second luminosity peak (SEc) is found but with no clear convergence peak associated with it. This bimodality of the central galaxy distribution was mentioned in \cite{bonafede2014}. In their analysis, it was suggested that NWc and SEc could be two subclusters undergoing a major merger. Since our central galaxy peaks are consistent with theirs, we have adopted their nomenclature. In addition, \cite{bonafede2014} found a galaxy overdensity in the south-east, coinciding with the SE radio relic. Our SE convergence peak is roughly in the same location as the galaxy overdensity (coined SEext) from \cite{bonafede2014} and is statistically (see the bootstrap analysis below) consistent with our SE luminosity peak. In Section \ref{section:relating}, we discuss the statistical significance of any offset between the mass and galaxies. The NW convergence peak has a weak counterpart in galaxy luminosity.  

One way to test the significance of each structure that is detected in the convergence map is bootstrapping. We perform bootstrapping by re-sampling the source catalog with replacement 1,000 times. The bootstrapped catalogs are then processed through the same method to produce 1,000 convergence maps. From the sample of 1,000 bootstraps, we calculate the rms noise of the data and divide the original convergence map by the noise map to find the S/N ratio. The S/N in Figure \ref{fig:signal_to_noise} is a measure of the statistical significance of the structures observed in the surface mass density map. The NWc peak of the mass distribution is detected at the $\mytilde9\sigma$ level. The SE and NW structures are less significant with a $\mytilde 3\sigma$ detection each. Bootstrapping only predicts the statistical significance of the weak-lensing signal. Additional uncertainty from systematics could be present.

\subsubsection{Hubble Space Telescope Convergence Map}
\begin{figure*}[!htb]
    \centering
    \includegraphics[width=\textwidth]{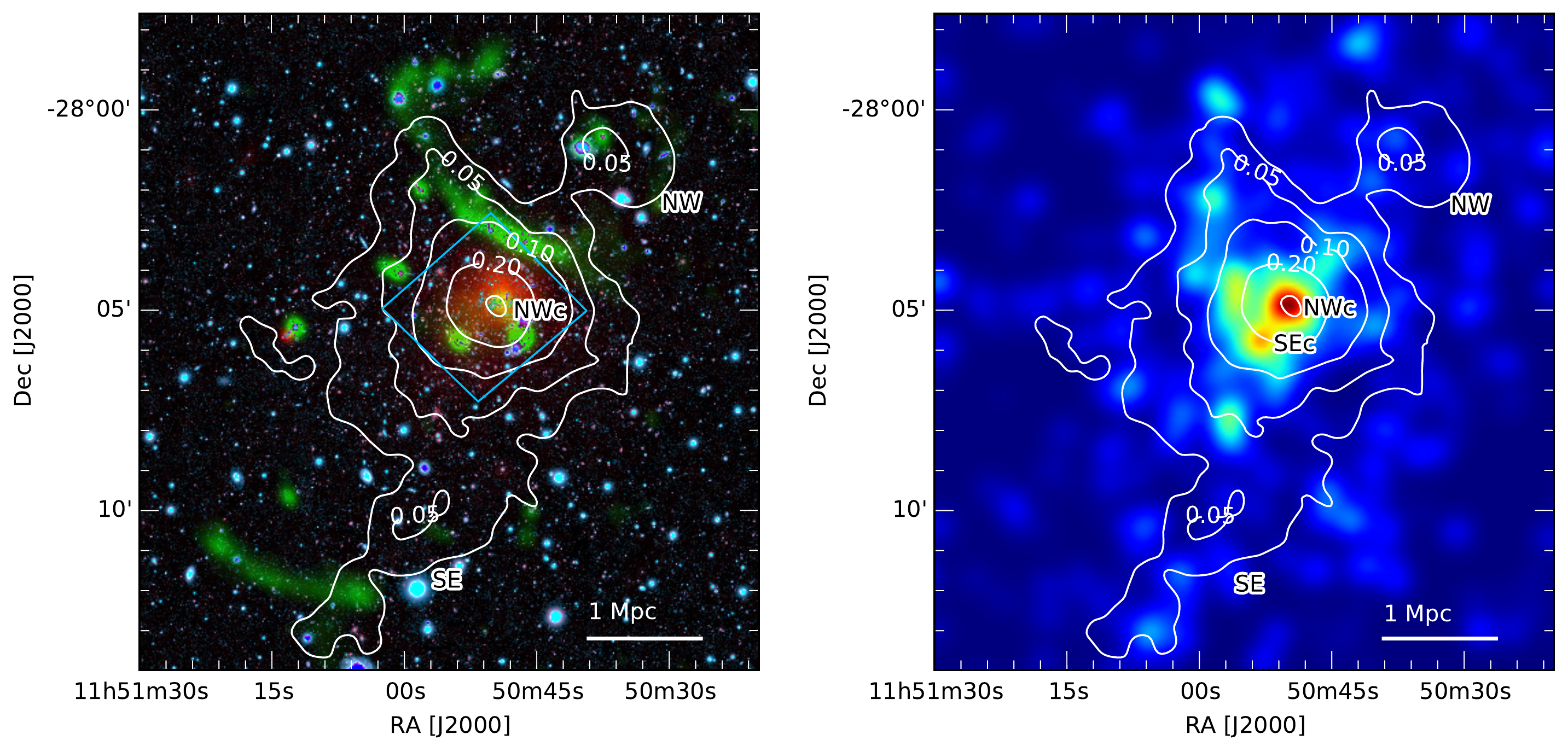}
    \caption{\label{fig:subaru_convergence} \textit{Left:} Color-composite Subaru image with enhanced radio (green) and X-ray (red) emissions. Radio emissions are from GRMT observations \citep{bonafede2014} and X-ray emissions are from \textit{XMM-Newton}. The overlaid convergence contours (white) peak in the X-ray emitting ICM with the highest contour enveloping the BCG. The mass is distributed along a direction similar to the axis connecting the radio relics. The \textit{HST} F814W pointing is outlined in light blue. \textit{Right:} Mass distribution overlaid on the Subaru cluster member catalog luminosity density. The luminosity density has a bimodal distribution in the cluster center with the NWc luminosity peak consistent with the mass peak. The SEc luminosity peak has no clear mass counterpart in the Subaru mass reconstruction. }
\end{figure*}

\begin{figure*}[!htb]
    \centering
    \includegraphics[width=0.5\textwidth]{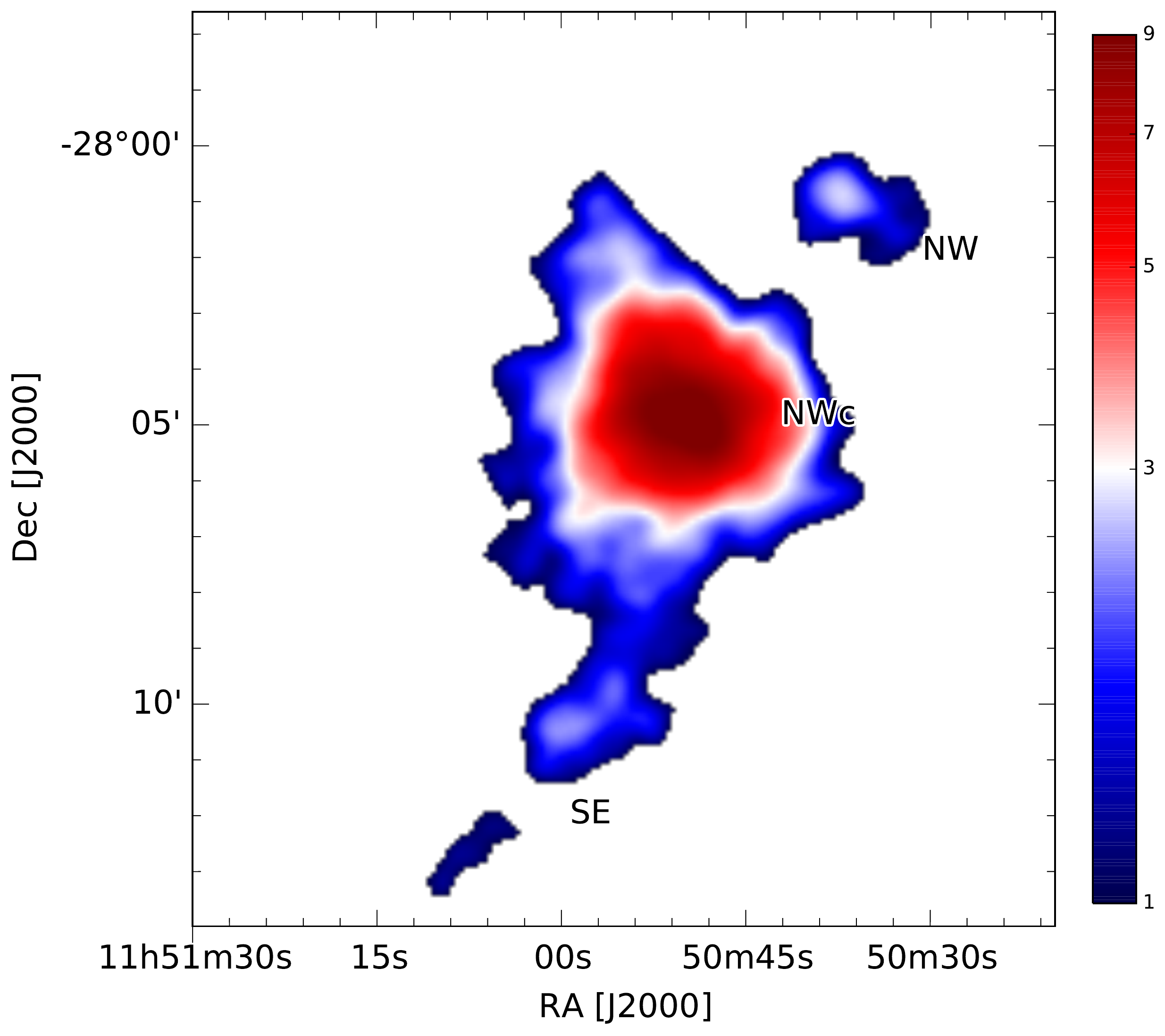}
    \caption{\label{fig:signal_to_noise} Subaru S/N map obtained by 1,000 bootstraps of the Subaru source catalog. The signal is the original convergence map and the noise is determined from the standard deviation of the bootstrapped convergence maps. The peak of the mass distribution is detected at $9\sigma$ significance and the substructures to the north-west and south-east are detected at $3\sigma$ significance.}
\end{figure*}

To probe the cluster at a higher resolution, we make use of deep \textit{HST} imaging. The observation footprint of the \textit{HST} imaging (blue box in Figure \ref{fig:subaru_convergence}) is much smaller than the Subaru observations, covering the central $3\arcmin \times 3\arcmin$ ($\mytilde 1\ \text{Mpc} \times 1\ \text{Mpc}$) field of the cluster. The average distortion within this subfield is high because the region is within a strong-lensing regime. This high distortion accompanied by the high source density ($\mytilde 100\ \mbox{arcmin}^{-2}$) allows us to use a smaller kernel to average the ellipticities for mass reconstruction. We choose a $20\arcsec$ radius top-hat kernel to average ellipticities for the \textit{HST} source catalog. The expected S/N within this $20\arcsec$ kernel is comparable to the one for the $80\arcsec$ kernel applied to the Subaru catalog.

\begin{figure*}[!htb]
\centering
\includegraphics[width=\textwidth]{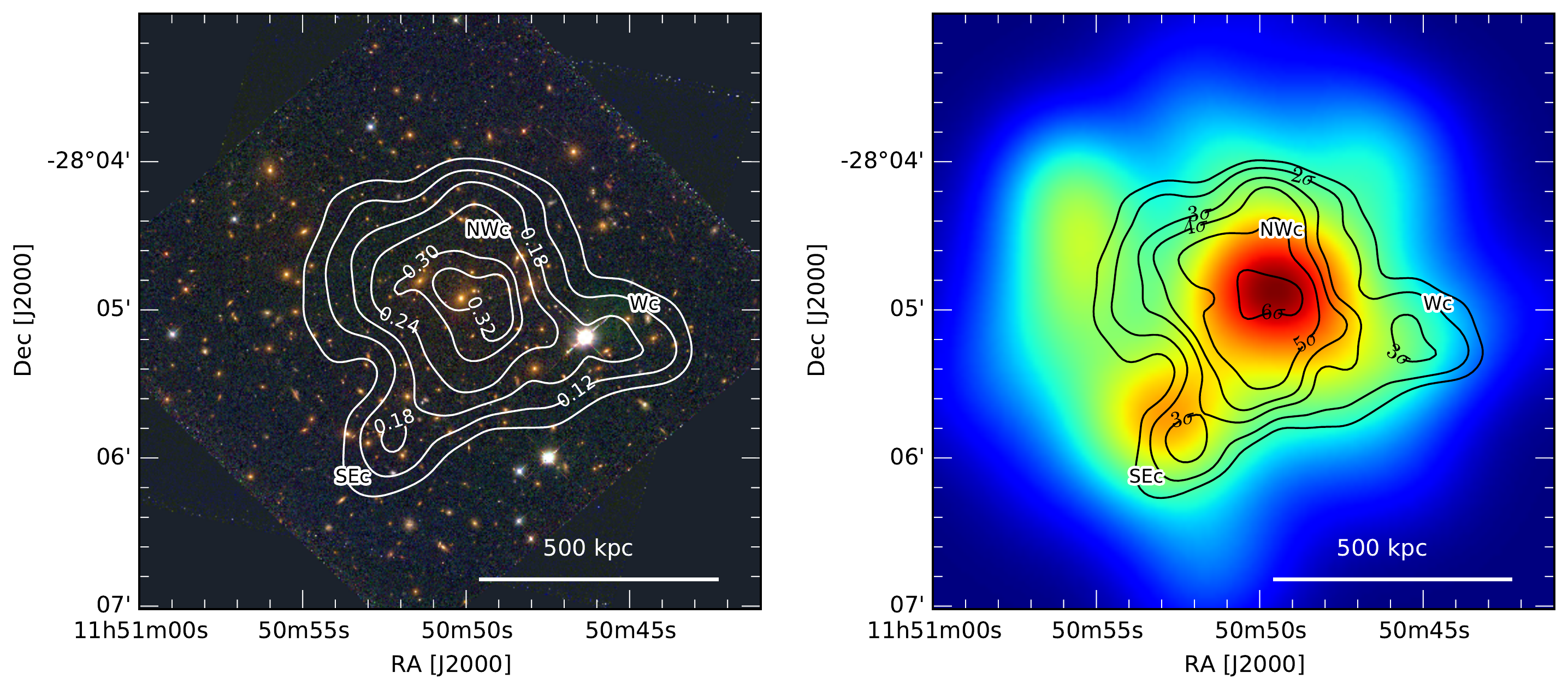}
\caption{\textit{Left:} \textit{HST} convergence overlaid on \textit{HST} color-composite image. The color-composite \textit{HST} image is created using F435W+F475W, F606W, and F814W to represent intensities in blue, green, and red, respectively. The BCG is located within the highest convergence contour. The high resolution of the \textit{HST} convergence map allows the detection of the SEc substructure $\mytilde 400$ kpc south-east of the BCG. Also of note is the extension of the convergence to the west. \textit{Right:} S/N contours plotted over the luminosity density of the \textit{HST} cluster member catalog. The primary peak is detected at the $6\sigma$ level. The SEc luminosity peak has a corresponding convergence peak but an offset is visible. The western convergence peak has a weak luminosity counterpart.}
\label{fig:hst_convergence}
\end{figure*}

The left panel of Figure \ref{fig:hst_convergence} shows the \textit{HST} convergence map overlaid on an \textit{HST} color composite. It is remarkable that the primary convergence peak that was detected in the Subaru reconstruction as a single peak is now resolved into three peaks in the \textit{HST} reconstruction. This weak-lensing mapping of the convergence has a striking resemblance to the strong-lensing map of \cite{zitrin2017}. The primary peak (NWc) is consistent with the BCG, a secondary peak (SEc) is observed $\mytilde400$ kpc to the south-east and a substructure is found to the west (Wc). 

We test the significance of the substructures by bootstrapping. The resulting \textit{HST} S/N map is shown in the right panel of Figure \ref{fig:hst_convergence}. The color map is the smoothed luminosity distribution of cluster member galaxies. The primary peak, NWc, is detected at the $6\sigma$ level and lies on top of the luminosity peak. A $3\sigma$ detection of the SEc mass structure is made. An offset ($\mytilde12\arcsec$) from the luminosity peak is observed and is further analyzed in Section \ref{section:relating}. Wc is detected with a statistical significance of $3\sigma$.

In summary, our Subaru and \textit{HST} mass reconstructions of PLCK~G287.0+32.9 find five structures that are detected at $\ge3\sigma$ statistical significance. The NWc peak (1) is the dominant feature in both the Subaru and \textit{HST} convergence maps. An analysis of \textit{HST} imaging shows that the cluster has two additional substructures detected in close proximity to the NWc peak. The SEc peak (2) is a substructure $\mytilde1\arcmin$ south-east of NWc and has an optical counterpart. The other structure lies to the west, thus named the Wc peak (3), and has a weak optical detection of galaxy light. Looking at the large scale mass distribution provided by the Subaru convergence map, a NW peak (4) is to the north-west beyond the NW radio relic and a SE peak (5) is located in the south-east towards the SE radio relic. In the following sections, the mass of each substructure will be estimated and their relation to the cluster galaxies and ICM will be analyzed.

\subsection{Substructure Analysis} \label{section:convergence peaks}
The detection of multiple peaks in the mass maps reveal PLCK~G287.0+32.9 is a complex system. However, mass reconstruction is a noisy inversion procedure. It is possible that spurious substructures arise because of chance alignments of background galaxies. Here, we quantify the significance of the substructures and their centroids. 

\begin{figure*}[ht!b]
\centering
\includegraphics[width=\textwidth]{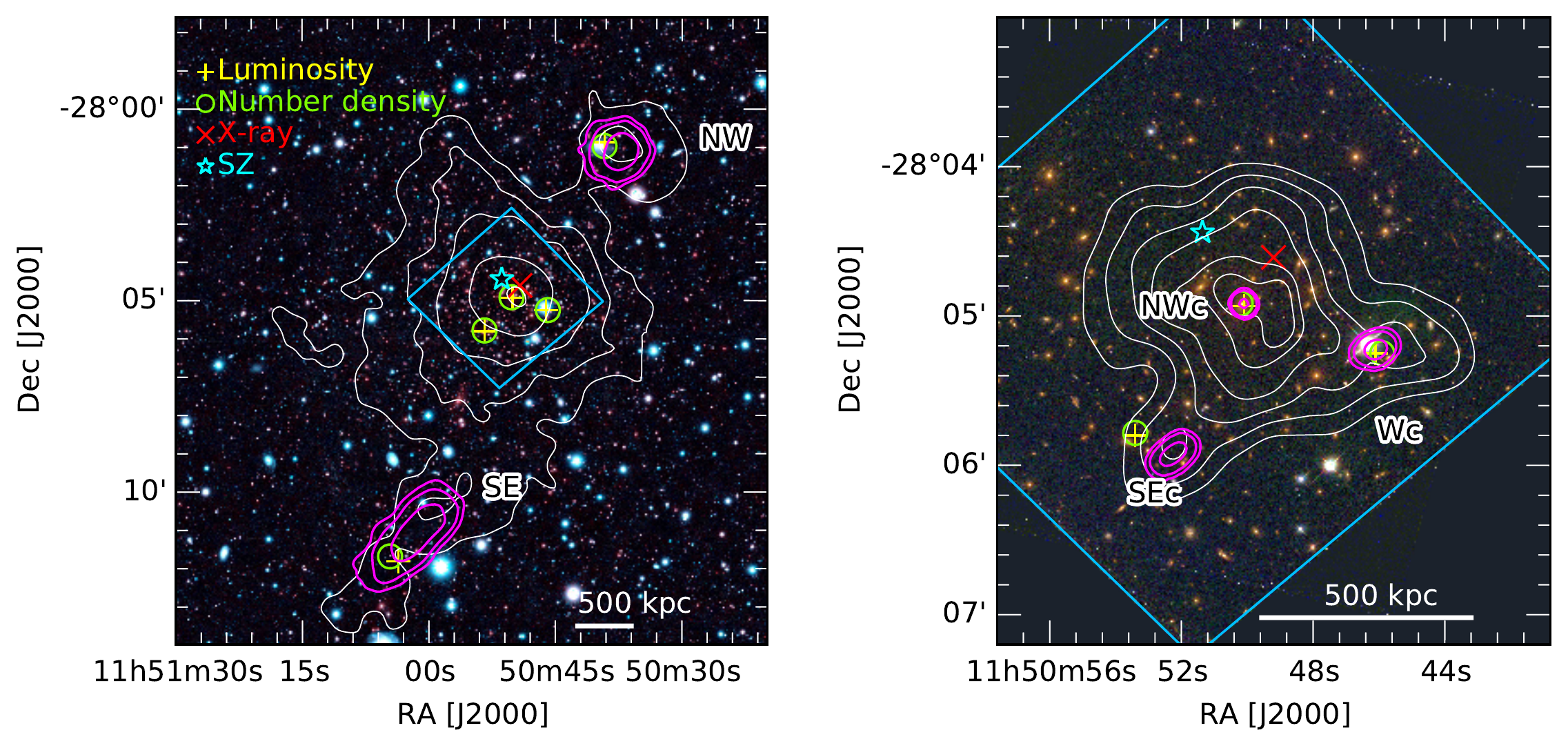}
\caption{Convergence peak distributions (magenta contours) determined from the bootstrapped convergence realizations. Convergence peaks are found within top-hat priors that are centered on the local luminosity centroid. The peaks are smoothed using a Gaussian kernel density estimator and the $1\sigma$, $2\sigma$, and $3\sigma$ contours are plotted. Galaxy luminosity and density centroids are shown, as well as X-ray and SZ peaks from \cite{bagchi2011}. \textit{Left:} Peak distributions of the NW and SE convergence are found using the Subaru bootstraps. Subaru weak lensing mass contours (white) are shown for reference. \textit{Right:} Peak distributions of the NWc, SEc, and Wc peaks from \textit{HST} bootstraps. The \textit{HST} peak distributions may be influenced by the tight priors that are required to prevent influence from neighboring peaks.}
\label{fig:color_peaks}
\end{figure*}

We use the galaxy distribution as a supplement to the convergence map to identify substructures. By looking at spatial correlations between the galaxy distribution and convergence peaks, we test the five substructures. For each substructure, we calculate the luminosity centroid by its luminosity-weighted moment within a top-hat prior. The galaxy number centroid is estimated in a similar fashion except that each galaxy is treated equally. The top-hat prior is centered on the brightest, confirmed cluster member nearest the convergence peak and the radius of the top-hat function depends on the distance to the nearest substructure. Table \ref{table:priors} shows the positions and radii of the top-hat windows. The resulting positions of the luminosity and number density centroids are labeled in Figure \ref{fig:color_peaks} with plus (yellow) and circle (green) symbols, respectively.

\begin{table*}[ht]
\centering
\caption{Top-hat Prior Parameters}
\makebox{
\def\arraystretch{1.5}
\begin{tabular}{lccccc}

\hline \hline
Name                     & R.A. & Dec. & Radius  \\
& & & (arcsec) \\
\hline
NWc        & 11$^\text{h}$50$^\text{m}$50$^\text{s}$.1 & -28$^\circ$04\arcmin56\arcsec & 35 \\
SEc        & 11$^\text{h}$50$^\text{m}$53$^\text{s}$.6 & -28$^\circ$05\arcmin50\arcsec & 35 \\
Wc        & 11$^\text{h}$50$^\text{m}$46$^\text{s}$.7 & -28$^\circ$05\arcmin17\arcsec & 35 \\
NW        & 11$^\text{h}$50$^\text{m}$39$^\text{s}$.4 & -28$^\circ$00\arcmin42\arcsec & 125 \\
SE        & 11$^\text{h}$51$^\text{m}$03$^\text{s}$.6 & -28$^\circ$12\arcmin60\arcsec & 200 \\
\hline

\hline
\end{tabular}
}
\label{table:priors}
\end{table*}

We utilize the 1,000 bootstrap results of the Subaru and \textit{HST} convergence maps to investigate the statistical significance of the galaxy and convergence alignment. The Subaru data are used to find the peak distribution of the NW and SE substructures. We choose not to analyze the NWc peak in the Subaru bootstraps because it is too close to the SEc and Wc peaks that are not resolved in the Subaru convergence. The NWc and SEc peak distributions are determined from the \textit{HST} bootstraps. The 1,000 peak centroids are smoothed by a Gaussian kernel density estimator in order to create the probability density map. Figure \ref{fig:color_peaks} shows the 1$\sigma$, 2$\sigma$, and 3$\sigma$ contours of each convergence peak distribution. One caveat is that the priors used to find the \textit{HST} convergence peak distributions are very restricted due to the close proximity of the NWc and SEc peaks. Therefore, it is possible that the significance contours are artificially tightened by this restrictive prior. We find that the NWc, Wc, NW, and SE substructures have convergence peak distributions that are within $2\sigma$ of the galaxy centroids. The fifth, SEc, is outside $2\sigma$ but we warrant its inclusion in our mass analysis because of the nearby luminosity peak, the limited size of the window due to the close proximity of NWc, and the detection of this structure in the strong-lensing analysis of \cite{zitrin2017}. We defer the detailed discussion of the relation of the galaxy distribution and convergence map to Section \ref{section:relating}. 

\subsection{Mass Estimation} \label{section:mass estimation}
A traditional approach in quantifying a cluster mass is to approximate the entire mass distribution with a single profile based on a tangential shear profile. However, this approximation becomes inadequate and leads to bias for merging galaxy clusters, which possess multiple halos. Nevertheless, as we have shown in Section \ref{section:mass density}, PLCK~G287.0+32.9 is dominated by the primary mass peak coincident with the cluster BCG. Therefore, in this study, we choose to use this traditional method to estimate the total cluster mass and also present a comprehensive mass analysis accounting for the substructures. Careful estimation of the substructure masses is critical in our reconstruction of the cluster merger scenario and also provides valuable inputs to our future numerical simulations.

For our tangential shear calculations, we choose to use the BCG as the origin because both the Subaru and \textit{HST} mass reconstructions show that the BCG and the primary convergence peak are highly consistent. 
The reduced tangential shear and cross shear are calculated as: 
\begin{equation}
\begin{split}
 g_T &= -g_1 \cos2\phi - g_2 \sin2\phi ,\\
 g_{\times} &= g_1 \sin2\phi - g_2 \cos2\phi ,
\end{split}
\end{equation}
where $g$ is the ellipticity of source galaxies (as explained in Section \ref{section:lensing theory}). The angle $\phi$ is measured from the horizontal axis that intersects the defined cluster center to each galaxy position. The cross shear (or B-mode) is a $\pi / 4$ rotation of the galaxy ellipticities and should be devoid of signal. Significant deviations of the cross shear from zero are often attributed to residual systematic errors. We measure the tangential shear in annular bins of $\delta r=100\arcsec$. Weak-lensing approximations become invalid in the central region. Thus, galaxies within $R_{\text{min}}=100\arcsec$ of the center are omitted from the tangential shear fitting. This minimum radius is approximately twice the estimated Einstein radius ($\theta_E \approx 42 \arcsec$; \citealt{zitrin2017}).

\begin{figure}[!htb]
    \centering
    \includegraphics[width=0.5\textwidth]{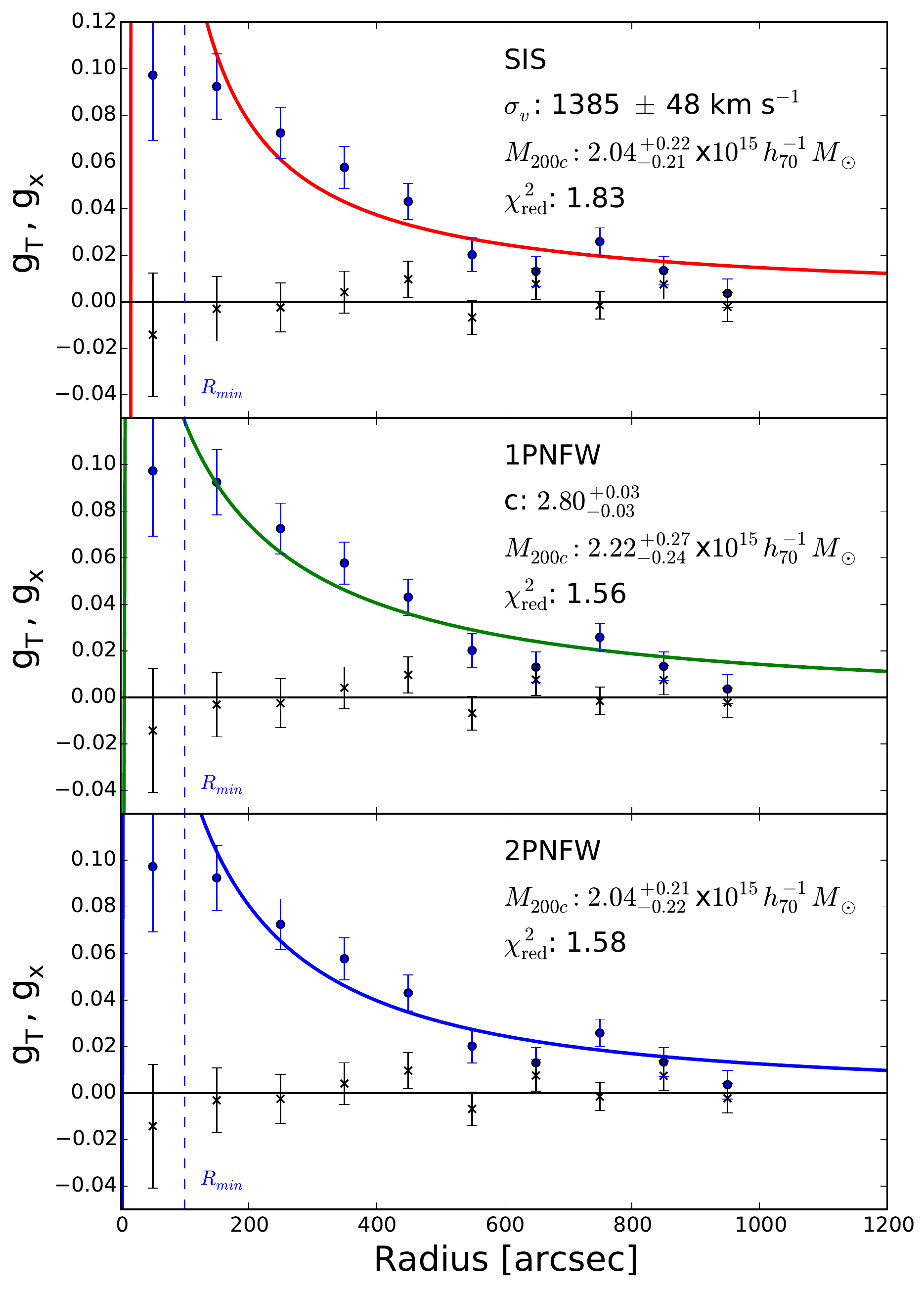}
    \caption{\label{fig:tan_nfw_mass} SIS and NFW halo fits to the reduced tangential shear. The radial profile is measured from the BCG. High S/N extending to large radii permits a 2-parameter NFW profile fit. Galaxies within $100\arcsec$ are excluded from the fitting algorithm to omit the central area where the weak lensing approximation may break down.} 
\end{figure}

Figure \ref{fig:tan_nfw_mass} shows the reduced tangential shear (blue dots) and cross shear (black x's) measured from the Subaru shape catalog. The cross shear is consistent with zero, suggesting minimal systematic errors. The tangential shear is highest near the cluster center, reaching $g_T=0.1$ and decreases with radius. The tangential shear is measured out to $1,000\arcsec$ ($\mytilde5$~Mpc), beyond which an annulus does not complete a circle.

The mass of the cluster is estimated by fitting a density profile to the reduced tangential shear. 
We note that, in this fit, we assume only shape noise in our covariance, i.e. the scatter due to the intrinsic shapes of background galaxies. At the data quality of Subaru and HST data, this is an important but not the only relevant source of mass uncertainty. Uncorrelated large-scale structure along the line of sight to the cluster \citep[e.g., ][]{Hoekstra2003}, and variations of cluster mass profiles at fixed mass \citep[e.g.,][]{Becker2011} cause an uncertainty floor of $\approx20$ per cent \citep[e.g.,][]{Gruen2015} for massive clusters. The uncertainty on mass due to shape noise ($\approx 10$ per cent) that we quote are thus not the full error on mass. Likewise, we do not expect the residuals of the fit, weighted by the inverse shape noise covariance, to follow a $\chi^2$ distribution, but rather to be somewhat larger.

A simple density profile is the singular isothermal sphere (SIS). The reduced shear for an SIS model is:

\begin{equation}
g = \frac{\gamma}{1-\kappa}=\frac{1}{(2x/\theta_E)-1} , 
\end{equation}
where the projected radius is $x$ and the Einstein radius is:
\begin{equation}
\theta_E = 4\pi \left(\frac{\sigma_v}{c}\right)^2\frac{D_{ls}}{D_s},
\end{equation}
where $\sigma_v$ is the velocity dispersion and $c$ is the speed of light. In the top panel of Figure \ref{fig:tan_nfw_mass}, the red curve shows the SIS model fit to the reduced tangential shear by least-squares minimization. The model returns the velocity dispersion $\sigma_v=1385\pm48\ \text{km}\ \text{s}^{-1}$. This velocity dispersion is much lower than $\sigma_v=1697\pm87\ \text{km}\ \text{s}^{-1}$ derived from the redshift distribution in Section \ref{section:spectroscopy}. The difference reinforces the importance of an equilibrium independent mass estimation when considering a merging galaxy cluster. The radius where the SIS model density is 200 times the mean critical density of the Universe is $r_{200\text{c}} = 2.29\pm0.08$ Mpc (at $z=0.385$). The mass within $r_{200\text{c}}$ is $M_{200\text{c}}=2.04^{+0.22}_{-0.21} \times 10^{15}\ h^{-1}_{70}\ \text{M}_{\odot}$.

\begin{figure}[!htb]
    \centering
    \includegraphics[width=0.5\textwidth]{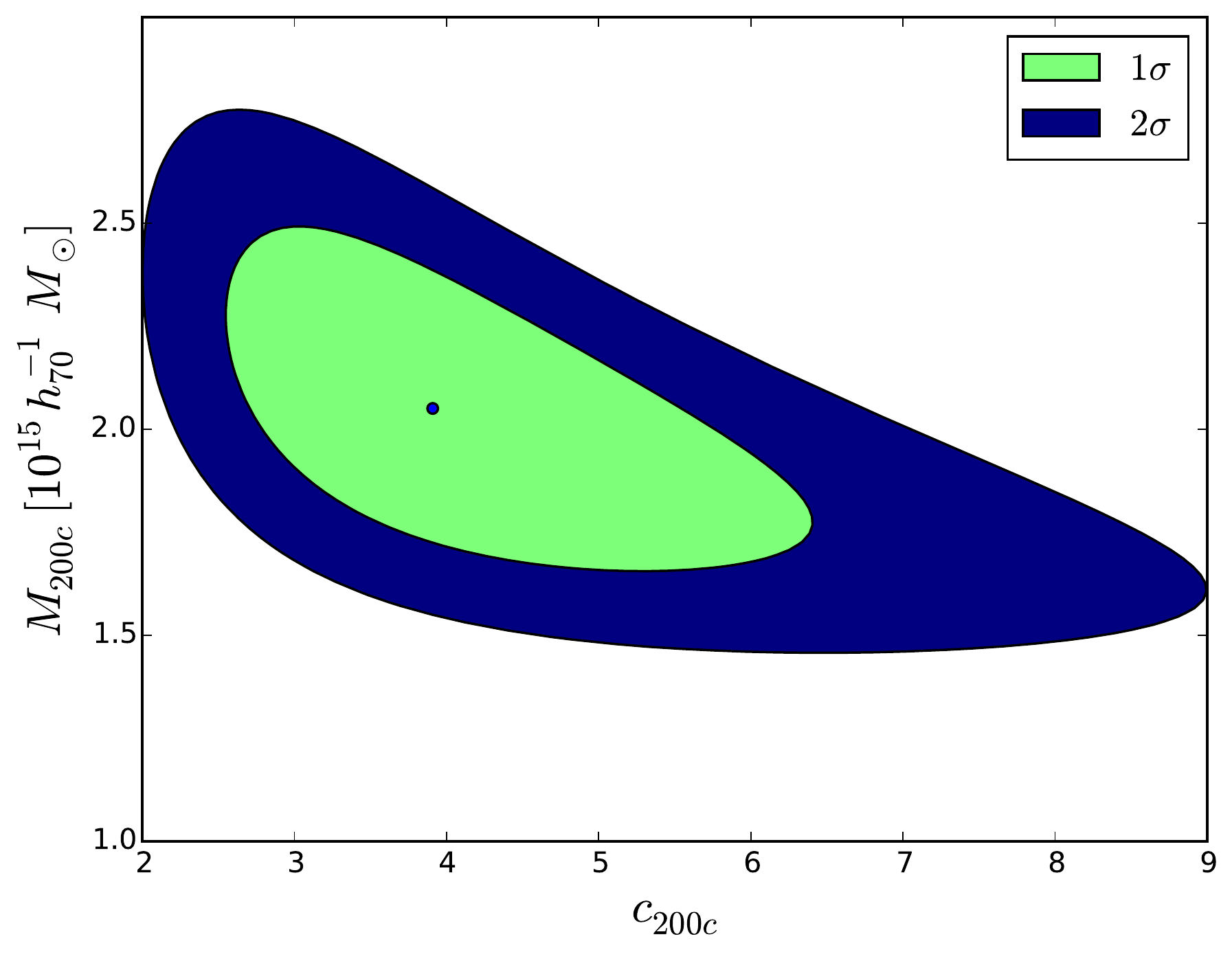}
    \caption{The mass-concentration parameter space of the 2PNFW fit to the tangential shear. Blue signifies the $1\sigma$ and green the $2\sigma$ uncertainties on the parameters. Each parameter is marginalized over to find the uncertainty of the 2PNFW fit.}
    \label{fig:nfw_uncertainty}
\end{figure}
 
A better description of a dark matter halo is the NFW profile. The density profile of an NFW halo is described by two parameters: the scale radius ($r_s$) and the dimensionless concentration ($c$). Compared to the SIS model, the NFW profile is steeper at large radii and shallower at small radii. We use the mathematical formulation as described in \cite{wright2000gravitational} to model the NFW reduced tangential shear. In general, because of the degeneracy between $r_s$ and $c$, a mass-concentration relation is assumed when fitting an NFW profile. However, the extreme mass of the cluster and the high S/N of the lensing signal over a large range permits fitting an NFW profile without relying on a mass-concentration relation. Therefore, our main mass estimate is obtained without any mass-concentration assumption (hereafter 2PNFW). To compare with previous studies, we also provide masses estimated using the mass-concentration relation of \cite{duffy2008dark} (hereafter 1PNFW).   

The 2PNFW approach gives $M_{200\text{c}} = 2.04^{+0.21}_{-0.22} \times 10^{15}\ h^{-1}_{70}\ \text{M}_{\odot}$ and $r_{200\text{c}}=2.29\ \text{Mpc}\ (435\arcsec)$. The best fit parameters are centered at ($c_{200\text{c}}$, $r_s$) = (3.92, $111\arcsec$). As mentioned above, these two parameters are highly degenerate. The uncertainty of the 2PNFW mass is estimated by an MCMC of the parameter space. Figure \ref{fig:nfw_uncertainty} gives the $1\sigma$ (green) and $2\sigma$ (blue) filled contours for the uncertainty of the mass-concentration relation. The contours highlight the degeneracy of $r_s$ and $c$ but a tight relation is seen along the mass axis.

The middle panel of Figure \ref{fig:tan_nfw_mass} shows that the 1PNFW profile fits the reduced tangential shear well. The estimated mass of the cluster is $M_{200\text{c}}=2.22^{+0.27}_{-0.24} \times 10^{15}\ h^{-1}_{70}\ \text{M}_{\odot}$ with $c_{200\text{c}}=2.80^{+0.03}_{-0.03}$ and $r_{200\text{c}} = 2.35\pm0.09$ Mpc. The mass-concentration relation is determined from simulations using WMAP5 cosmology and the scatter in the relation introduces additional uncertainty when used to derive mass.

\subsection{Multiple Halo Fit} \label{section:multi-halo fit}
Radio relics, a radio halo, and a disturbed X-ray morphology suggest that PLCK~G287.0+32.9 is a merging galaxy cluster. Adding to this evidence, we detect a complex mass distribution and define five structures. Therefore, modeling the cluster as a single halo may bias the mass estimation. A more proper method is to use a superposition of multiple halos to model the lensing signal. 

For our multiple halo fitting a combined Subaru and \textit{HST} source catalog is created by joining the catalogs with the \textit{HST} shapes taking precedence for all spatially matched sources. Multiple halo fitting is done by minimizing the difference between the measured ellipticity and a modeled ellipticity for each source galaxy. We assume each substructure follows an NFW halo with the same mass-concentration relation as above and model the expected ellipticity (shear) at each source galaxy position. The 2PNFW model is not used because the added freedom leads to divergence when fitting. Model ellipticities for each galaxy are constructed while considering multiple halos simultaneously. Each halo is given an initial concentration and is fixed to its convergence peak. Then, the shear at each galaxy position is modeled by considering the contribution from all halos. Halos are fixed at $z=0.385$ with $\beta=0.523$ and width of the redshift distribution, $\langle\beta^2\rangle / \langle\beta \rangle^2 - 1=0.218$ for galaxies from the Subaru source catalog and $\beta=0.545$, $\langle\beta^2\rangle / \langle\beta \rangle^2 - 1=0.163$ for the \textit{HST} galaxies. The contribution of each galaxy to the complete model is weighted by the uncertainty of its ellipticity as
\begin{equation}
\sigma_i = \frac{1}{\sqrt{\delta e_i^2 + \sigma_{SN}^2}}, 
\end{equation}
where $\delta e_i$ is the ellipticity uncertainty of the $i^{th}$ galaxy and $\sigma_{SN}$ is the shape noise ($\mytilde0.25$).

\begin{table*}[ht]
\centering
\caption{Single halo fitting NFW parameters.}
\makebox{
\def\arraystretch{1.5}
\begin{tabular}{lccccc}

\hline \hline
Peak                     & R.A. & Dec. & $c_{200\text{c}}$ & $M_{200\text{c}}$\tablenotemark{a}  & $r_{200\text{c}}$ \\
& & & & ($10^{14} \ h^{-1}_{70}\ \text{M}_{\odot}$) & Mpc \\
\hline
NWc        & 11$^\text{h}$50$^\text{m}$50$^\text{s}$.3 & -28$^\circ$04\arcmin52\arcsec & $2.79\pm0.02$ & $23.1^{+1.7}_{-1.7}$ & $2.38\pm0.06$\\
X-ray      & 11$^\text{h}$50$^\text{m}$49$^\text{s}$.2 & -28$^\circ$04\arcmin37\arcsec & $2.82\pm0.02$ & $20.7^{+1.7}_{-1.5}$ & $2.30\pm0.06$\\
SZ         & 11$^\text{h}$50$^\text{m}$51$^\text{s}$.4 & -28$^\circ$04\arcmin26\arcsec & $2.84\pm0.02$ & $19.0^{+1.5}_{-1.5}$ & $2.23\pm0.06$\\
\hline

\hline
\end{tabular}

}
\tablenotetext{a}{Quoted errors are from shape noise only}

\label{table:single_halo}
\end{table*}

\begin{table*}[ht]
\centering
\caption{Five halo fitting NFW parameters.}
\makebox{
\def\arraystretch{1.5}
\begin{tabular}{lccccc}

\hline \hline
Peak                     & R.A. & Dec. & $c_{200\text{c}}$ & $M_{200\text{c}}$\tablenotemark{a}  & $r_{200\text{c}}$ \\
& & & & ($10^{14} \ h^{-1}_{70}\ \text{M}_{\odot}$) & Mpc \\

\hline
NWc     & 11$^\text{h}$50$^\text{m}$50$^\text{s}$.3 & -28$^\circ$04\arcmin52\arcsec & $2.88\pm0.03$ & $15.9^{+2.5}_{-2.2}$ & $2.10\pm0.10$ \\
SEc     & 11$^\text{h}$50$^\text{m}$51$^\text{s}$.7 & -28$^\circ$05\arcmin50\arcsec & $3.59\pm0.16$ & $1.16^{+0.15}_{-0.13}$ & $0.88\pm0.03$ \\
Wc      & 11$^\text{h}$50$^\text{m}$45$^\text{s}$.0   & -28$^\circ$05\arcmin04\arcsec & $3.80\pm0.21$ & $0.59^{+0.06}_{-0.06}$ & $0.70\pm0.03$ \\
NW      & 11$^\text{h}$50$^\text{m}$37$^\text{s}$.0 & -28$^\circ$01\arcmin00\arcsec & $3.45\pm0.11$ & $1.87^{+0.24}_{-0.22}$ & $1.03\pm0.04$ \\
SE      & 11$^\text{h}$50$^\text{m}$59$^\text{s}$.6 & -28$^\circ$10\arcmin07\arcsec & $3.48\pm0.11$ & $1.68^{+0.22}_{-0.20}$ & $1.00\pm0.04$ \\
\hline

\hline

\end{tabular}
\tablenotetext{a}{Quoted errors are from shape noise only}

}

\label{table:five_halo}
\end{table*}

We fit a single halo to compare this method with the tangential shear fit (Section \ref{section:mass estimation}). We fix the halo at the BCG, the same center used in tangential shear fitting, and allow the concentration to vary. Consistent with the tangential shear method, galaxies within $100\arcsec$ of the peak are excluded from the fitting algorithm. The result is summarized in Table \ref{table:single_halo}. The fitted concentration is $c_{200\text{c}}=2.79\pm0.02$, the mass of the halo is $M_{200\text{c}} = 2.31^{+0.17}_{-0.17}\times10^{15} \ h^{-1}_{70}\ \text{M}_{\odot}$, and $r_{200\text{c}}=2.38\pm0.06$ Mpc. As expected, fitting an NFW halo by modeling the contribution of each galaxy is consistent with that of tangential shear fitting. Table \ref{table:single_halo} also contains the masses derived based on the X-ray and SZ peaks, which are consistent with the mass estimated using the BCG as the center. 

A five halo model is fit to the five structures that were defined from the convergence maps. Halo models are built simultaneously with each centroid fixed to its convergence peak. The masses derived from fitting five halos are summarized in Table \ref{table:five_halo}. Multiple halo fitting shows that NWc is the primary mass of the system. SEc, NW, and SE are approximately a tenth of the mass of NWc based on the five halo fitting. The Wc structure is low mass and is insignificant compared to the others. Note that multiple halo fitting should be more robust than the 1D tangential shear fitting for estimating the mass of a merging galaxy cluster but the accuracy of this method is still limited by the imposed model.

\section{Discussion}
\subsection{Comparison with Other Mass Estimates}
As the second most significant detection from the Planck SZ survey, PLCK~G287.0+32.9 is expected to be extremely massive. This extreme mass is supported by the high X-ray luminosity ($L_{X}=17.20\pm0.11\times 10^{44}\ \text{erg}\ \text{s}^{-1}$) measured by \textit{XMM-Newton}. Also, the recent strong-lensing analysis by \cite{zitrin2017} has classified PLCK~G287.0+32.9 among the largest Einstein radius ($\theta_E\approx42\arcsec$ for source at $z\simeq2$) clusters detected. 

The mass of the cluster was first estimated by the $M_{500\text{c}}$ - $Y_{X,500\text{c}}$ relation to be $M_{500\text{c}} = 15.72\pm 0.27 \times10^{14}\ \text{M}_{\odot}$ \citep{planck2011b}. This method uses an equilibrium assumption where $Y_{X,500\text{c}}$ is the product of the gas mass inside $R_{500\text{c}}$ and the gas temperature measured by \textit{XMM-Newton} observations. However, mass estimates of merging galaxy clusters that rely on the hydrostatic equilibrium assumption are prone to large systematic errors due to the severe departure from the hypothesized equilibrium of the merging system. 

The first weak-lensing mass estimate was provided by \cite{gruen2014weak} with MPG/ESO telescope/Wide-Field Imager (WFI) observations. They measured the mass by fitting a single NFW halo to the shear centered at the BCG. Using the mass-concentration relation of \cite{duffy2008dark}, they estimated the mass to be $M_{200m}= 3.77^{+0.95}_{-0.76}\times 10^{15} \ h^{-1}_{70}\ \text{M}_{\odot}$ $(M_{200\text{c}}= 2.40^{+0.59}_{-0.48}\times 10^{15} \ h^{-1}_{70}\ \text{M}_{\odot})$.

In Section \ref{section:mass estimation}, we fit an NFW halo model to the reduced tangential shear without a mass-concentration relation. Our single-halo mass measurement based on reduced tangential shear fitting for PLCK~G287.0+32.9 is $M_{200\text{c}} = 2.04^{+0.21}_{-0.22} \times 10^{15} \ h^{-1}_{70}\ \text{M}_{\odot}$. Confining the mass-concentration relation to that of \cite{duffy2008dark}, we find the mass to be $M_{200\text{c}} = 2.22^{+0.27}_{-0.24} \times 10^{15} \ h^{-1}_{70}\ \text{M}_{\odot}$. This result is consistent with the result of \cite{gruen2014weak}. The tighter uncertainty ($\mytilde50$\% smaller) of our fit is a result of the improved statistics that arises from using deeper imaging. Fitting a single profile to a merging system has been shown to give a biased mass \citep[e.g.,][]{jee2015sausage}. A more careful method to measure the mass, presented in Section \ref{section:multi-halo fit}, gives $M_{200\text{c}} = 2.31^{+0.17}_{-0.17} \times 10^{15} \ h^{-1}_{70}\ \text{M}_{\odot}$.

\subsection{Comparing the Galaxy and Convergence Distributions} \label{section:relating}

Galaxy cluster merger scenarios can be tightly constrained by understanding the exact physical interaction among different cluster components: the dark matter mass distribution as reconstructed from weak lensing, the galaxy distribution observed in the optical, and the intracluster medium distribution observed by X-ray and radio emissions. In general, merging galaxy clusters have been shown to have coincident dark matter and galaxies while showing large offsets from the gas distribution. Conversely, some studies have shown dark matter coincident with gas and offset from the galaxy distribution \citep[e.g.,][]{mahdavi2007, jee2014a520}. Here we discuss the peak distributions for PLCK~G287.0+32.9 and its implications for the merging scenario.

\begin{figure*}[!htb]
\centering
\includegraphics[width=\textwidth]{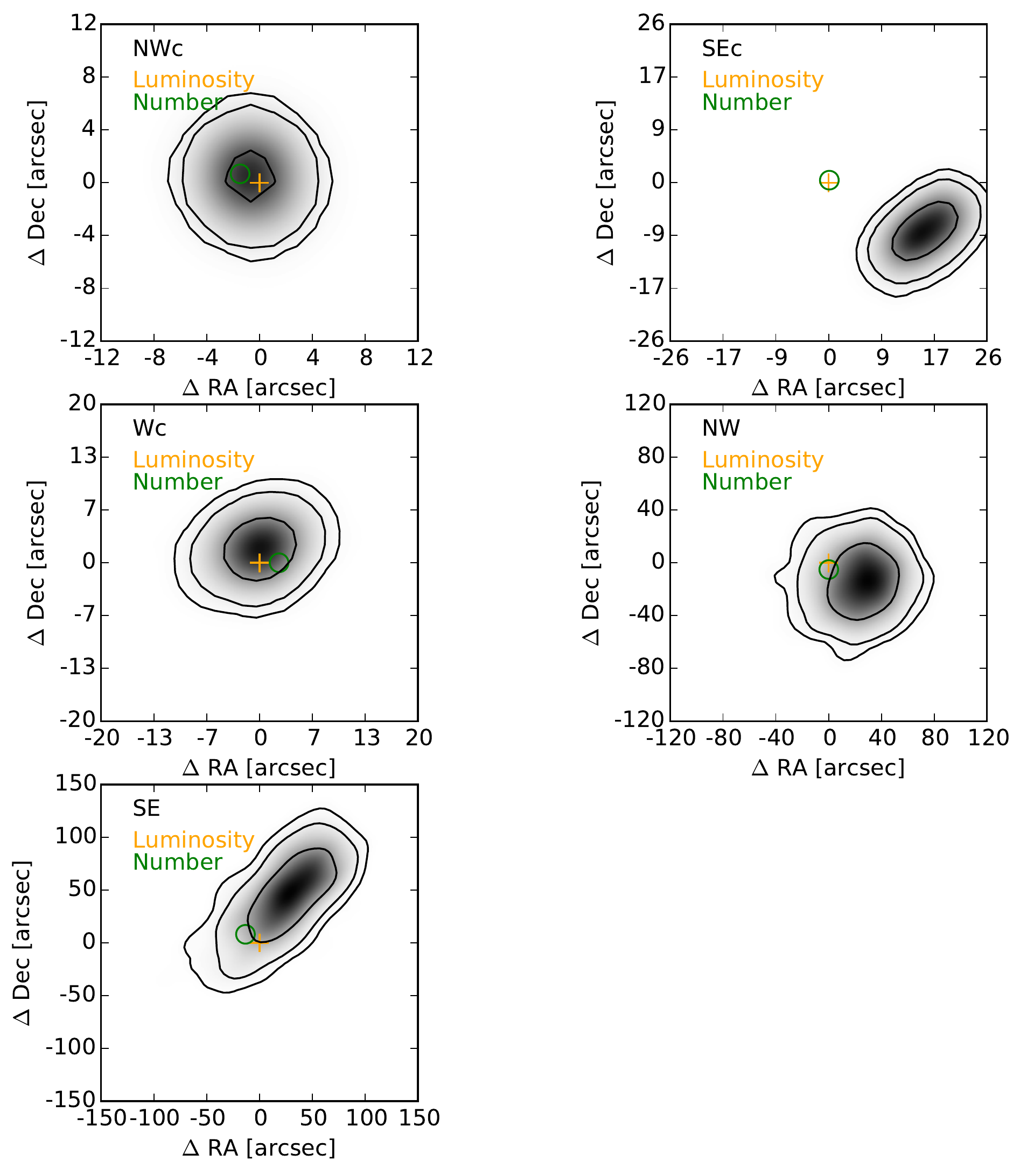}
\caption{Mass-galaxy offset significance for each of the substructures. Black contours show confidence levels of $1\sigma$, $2\sigma$, and $3\sigma$ for the mass peak distribution determined from bootstrapping. Plots are centered on the luminosity centroid (orange plus). Galaxy number density centroids are marked with a green circle.}  
\label{fig:peaks}
\end{figure*}

The mass peaks of each distribution are summarized in Table \ref{table:five_halo} and the distribution of mass peaks measured from bootstrap realizations of the mass map are plotted in Figure \ref{fig:color_peaks}. Figure \ref{fig:peaks} shows the relation between the mass peak distribution of each halo and the corresponding galaxy number density and luminosity peaks. The most massive clump, NWc (top left), shows a tight distribution of peaks and $1\sigma$ consistency with the galaxy peaks. On the other hand, the SEc (top right) mass distribution is offset from the galaxy peaks by $\mytilde 12\arcsec$ ($\mytilde 60$ kpc). This offset is also present in the convergence map presented in Figure \ref{fig:hst_convergence}. Measurement of the peaks in close proximity to the primary NWc cluster is a difficult task which requires a top-hat prior. Also, our weak-lensing assumption relating shears to ellipticity becomes invalid in the strong-lensing regime. The constraint of the top-hat prior may be constricting the distribution, though it did include the galaxy peaks. The NW and SE mass peak distributions are much broader, as expected from their low S/N detections (Figure \ref{fig:signal_to_noise}). Both the NW and SE mass peaks are 2$\sigma$ consistent with the galaxy peaks. Looking at the large scale distribution of the peaks, they are aligned with the axis connecting the radio relics, which is consistent with merger expectations.

\subsection{Mass Implications for the Merging Scenario}
An integral part of studying galaxy clusters is to provide mass constraints to be used in simulations. Galaxy clusters with ideal merger features, such as the Bullet \citep{clowe2006bullet} and Sausage \citep{jee2015sausage}, have been exhaustively analyzed from both observational and theoretical stances. PLCK~G287.0+32.9 is by no means as clear as these clusters appear to be, which makes it a good candidate for testing merger theories in a complex environment. For this reason, we provide results that may be used as constraints in future simulations of PLCK~G287.0+32.9. 

The geometry of the system is consistent throughout the data. Roughly, a merger axis can be defined by connecting the two radio relics through the cluster. Subaru optical observations show that the galaxy distribution is elongated in the south-east to north-west direction along the merger axis. The evidence presented by our weak-lensing analysis shows that the mass distribution on large scales also follows the same merger axis (Figure \ref{fig:subaru_convergence}). The higher resolution \textit{HST} mass map shows that two mass peaks (NWc and SEc) exist within the central mass clump and that these two mass peaks also lie along the merger axis.

One of the features that makes PLCK~G287.0+32.9 so unique is its asymmetric radio relics. The SE relic is located $\mytilde2.8$ Mpc from the X-ray peak, whereas the NW relic is only $\mytilde0.4$ Mpc. \cite{bonafede2014} proposed two merging scenarios that may result in the asymmetry of the relics. In one scenario, a low-mass group may have accreted from the north-west direction and created the two relics during two passes through the primary cluster. \cite{bonafede2014} mention that the remnant of this sub-group may be detected in the north-west. Our mass reconstruction provides locations of substructures that are helpful in constraining the merger scenario. The SEc and SE convergence substructures are consistent with the formation of the SE relic. In one scenario the relic is formed by a merger between the SE mass clump and the primary cluster. In order to explain the NW relic using this scenario, a second subcluster must have been involved. A second scenario could be the formation of both relics by two passes of the SEc substructure. In this scenario, the SE relic is created during first passage and the NW relic upon second passage. The Subaru mass map in Figure \ref{fig:subaru_convergence} shows a substructure in the NW that lies along the merging axis.  However, current theories predict that radio relics suffer little deceleration while subclusters do \citep{springel2007}. Because the NW substructure is beyond the radio relic, it is not likely the cause of the relic. 


\section{Conclusions}
A weak gravitational lensing analysis of PLCK~G287.0+32.9, the second most significant merging galaxy cluster of the Planck SZ survey, has been presented. The large mass of the cluster coupled with deep \textit{HST} and Subaru imaging allowed us to fit an NFW model, without the restriction of a mass-concentration relation, to the reduced tangential shear to confirm that the cluster is extremely massive, with mass of $M_{200\text{c}} = 2.04^{+0.20}_{-0.21}\times10^{15}\ \text{M}_{\odot}$. An eye catching feature of this cluster is its numerous radio emissions. Hosting two radio relics and a radio halo categorizes this cluster as merging. In this work, making use of new imaging, we were able to probe the mass distribution of the cluster and make the first discovery of five substructures. The mass is dominated by the primary cluster with three of the substructures being $\mytilde10\%$ of the mass of the primary cluster. The fifth substructure is low mass $\mytilde10^{13}$ and cannot be considered a galaxy cluster. Based on the spatial distribution of the substructures, we found that two of them are consistent with merger scenarios that may have created the radio relics. We believe there are two ways to proceed in the future. One is to improve the imaging of the cluster by fully covering it with deep \textit{HST} imaging. The sensitivity of the \textit{HST} would allow a more thorough look into the substructures that lie in the outskirts of the cluster. The other is to model the merger with simulations using the masses that have been estimated in our work. The asymmetry of the radio relics is rare and would be a good test for our understanding of galaxy cluster mergers.

The authors would like to thank Annalisa Bonafede for providing radio data. Support for the current \textit{HST} program was provided by
NASA through a grant from the Space Telescope Science
Institute, which is operated by the Association of Universities
for Research in Astronomy, Incorporated.
M.J.J. acknowledges support for the current research 
from the National Research Foundation of Korea under the program 2017R1A2B2004644 and 2017R1A4A1015178. D.W. and N.G. acknowledge support from the National Science Foundation under Grant No. (1518246).
Support for DG was provided by NASA through Einstein Postdoctoral
Fellowship grant number PF5-160138 awarded by the Chandra X-ray
Center, which is operated by the Smithsonian Astrophysical Observatory
for NASA under contract NAS8-03060.

\bibliographystyle{aasjournal.bst}
\bibliography{mybib}
\end{document}